\documentstyle[12pt,aaspp4,tighten]{article}
\lefthead{von Montigny et al.}
\righthead{Multi-wavelength Observations of 3C~273}

\font\rms=cmr10 at 10truept

\def\ar{$\alpha_r$}
\def\g{$\gamma$}
\def\10#1{10$^{#1}$}
\def\c10#1{$\cdot$10$^{#1}$}
\def\etal{et al. }
\def\Gg{$\Gamma_\gamma $}
\def\n{$\nu$}
\def\Fn{$F_{\nu}$}
\def\nFn{$\nu F_{\nu}$}
\def\s{$\sigma $}
\def\3c{3C~273 }
\def\affil#1#2{\par\noindent\refp2$^{#1}$ #2}
\def\refp2{\parshape=2 0truecm 15.5truecm  0.4truecm 15.1truecm}
\def\deg#1{$ #1^\circ$}
\def\ts{TS }
\def\chisq{$\chi^2$ }
\def\pcm2sec{${\rm photons\ cm^{-2}s^{-1}}$}
\def\um{$\mu{\rm m}$}
\def\+-{$\pm$}
  
\begin{document}
\title{Multi-wavelength Observations of \3c in 1993-1995}

\noindent
C.~von~Montigny$^{1,23,}$\altaffilmark{24}\altaffiltext{24}{present address: Landessternwarte Heidelberg-K\"onigstuhl, D-69117 Heidelberg, Germany},
H.~Aller$^2$,
M.~Aller$^2$,
F.~Bruhweiler$^3$,
W.~Collmar$^{4}$,
T.J.-L.~Courvoisier$^{8}$,
P.~G.~Edwards$^{19}$,
C.~E.~Fichtel$^1$,
A.~Fruscione$^5$,
G.~Ghisellini$^{21}$,
R.~C.~Hartman$^1$,
W.~N.~Johnson$^{20}$,
M.~Kafatos$^6$, 
T.~Kii$^{19}$,
D.~A.~Kniffen$^7$,
G.~G.~Lichti$^{4}$,
F.~Makino$^{19}$,
K.~Mannheim$^{22}$,
A.P.~Marscher$^{15}$,
B.~McBreen$^{16}$,
I.~McHardy$^{17}$,
J.~E.~Pesce$^{9}$,
M.~Pohl$^{4}$,
E.~Ramos$^6$,
W.~Reich$^{10}$,
E.I.~Robson$^{11}$, 
K.~Sasaki$^{19}$,
H.~Ter\"asranta$^{12}$,
M.~Tornikoski$^{12}$,
C.~M.~Urry$^{9}$,
E.~Valtaoja$^{12,13}$,
S.~Wagner$^{14}$,
T.~Weekes$^{18}$

{\hsize=16.5truecm
\vskip .3cm 
\parindent=0 pt \obeylines
\affil{1}{\rms NASA/Goddard Space Flight Center, Code 661, Greenbelt, MD 20771}
\affil{2}{\rms Astronomy Department, University of Michigan, Dennison Building, Ann Arbor, MI, 48109-1090}
\affil{3}{\rms Dept. of Physics, Catholic University of America, Washington, D.C. 20064}  
\affil{4}{\rms Max-Planck-Institut f\"ur extraterrestrische Physik, 85740 Garching, Germany}  
\affil{5}{\rms Harvard-Smithsonian Center for Astrophysics, 60 Garden Street, Cambridge, MA 02138}  
\affil{6}{\rms Center for Earth Observations and Space Research, CSI, George Mason University, Fairfax, VA 22030}
\affil{7}{\rms Hampden-Sydney College, P.O. Box 862, Hampden-Sydney, VA 23943}  
\affil{8}{\rms INTEGRAL Science Data Center, 16 Chemin d'Ecogia, CH-1290 Sauverny, Switzerland }  
\affil{9}{\rms Space Telescope Science Institute, 3700 San Martin Drive, Baltimore MD 21218}  
\affil{10}{\rms Max-Planck Institut f\"ur Radioastronomie, Bonn, Germany}  
\affil{11}{\rms Joint Astronomy Centre, 660 N. Aohoku Place, Hilo HI 96720}  
\affil{12}{\rms Mets\"ahovi Radio Research Station, FIN-02540 Kylmala, Finland}  
\affil{13}{\rms Tuorla Observatory, University of Turku, SF-21500, Piikki\"o, Finland}  
\affil{14}{\rms Landessternwarte Heidelberg-K\"onigstuhl, D-69117 Heidelberg, Germany}  
\affil{15}{\rms Department of Astronomy, Boston University, 725 Commonwealth Ave., Boston MA 02215}  
\affil{16}{\rms Physics Department, Univ. College, Belfield, Dublin 4, Ireland}  
\affil{17}{\rms Department of Physics, University of Southampton, Southampton S09 5NH, UK}  
\affil{18}{\rms Whipple Observatory, Harvard-Smithsonian Center for Astrophysics, Box 97, Amado AZ 85645}   
\affil{19}{\rms Inst. of Space \& Astronaut. Science, Yoshinodai 3-1-1, Sagamihara, Kanagawa 229, Japan}
\affil{20}{\rms E.O. Hulburt Center for Space Research, Code 7650, Naval Research Laboratory, Washington, DC 20375} 
\affil{21}{\rms Osservatorio di Brera-Merate, V. Bianchi, 46, Merate, Italy}
\affil{22}{\rms Universit\"ats-Sternwarte, Geismarlandstrasse 11, D-37803 G\"ottingen, Germany}  
\affil{23}{\rms NAS/NRC Resident Research Associate, NASA/GSFC}  
}


\begin{abstract}
We present the results of the multi-wavelength campaigns on \3c in
1993-1995.  During the observations in late 1993 this quasar showed an
increase of its flux for energies $\ge$ 100 MeV from about
2.1\c10{-7}\pcm2sec\ to approximately 5.6\c10{-7}\pcm2sec\ during a
radio outburst at 14.5, 22 and 37 GHz. However, no one-to-one
correlation of the \g -ray radiation with any frequency could be
found. The photon spectral index of the high energy spectrum changed
from \Gg\ = (3.20\+-0.54) to \Gg\ = (2.20\+-0.22) in the sense that
the spectrum flattened when the \g -ray flux increased. Fits of the
three most prominent models (synchrotron self-Comptonization, external
inverse Comptonization and the proton initiated cascade model) for the
explanation of the high \g -ray emission of active galactic nuclei
were performed to the multi-wavelength spectrum of \3c. All three
models are able to represent the basic features of the
multi-wavelength spectrum. Although there are some differences the
data are still not decisive enough to discriminate between the models.
\end{abstract}

\section{Introduction}

The identification of 54 EGRET \g -ray sources with radio-loud active galactic
nuclei (AGN) (\cite{tho95}, 1996) has drawn the attention of the astronomical
community to these very interesting extragalactic objects.  
The observations have shown that for all AGN detected by EGRET the
energy output in the high energy regime is at least
as high as, and during  active periods can be as much as 10-30 times, that in
the optical regime.  An extensive summary of the results from phases~1 and 2
(1991 May 16 to 1993 August 17) of the observations of these AGNs with EGRET
and their implications has been given by von Montigny \etal (1995).

\3c shows all the characteristics which are typical for high luminosity
quasars: a flat radio spectrum of the core, strong and rapid
variability in the optical and other energy ranges (\cite{cou88},
\cite{cou90}), variable polarization, a radio jet with superluminal
motion.  Additionally, it shows an optical and X-ray jet and a very
prominent UV excess, the so-called ``big blue bump''. Although \3c is
not a high polarization quasar (HPQ) --- for which the standard
definition is that the {\it optical} polarization exceeds 3\% at least
once --- it is an interesting borderline case between high and low
polarization quasars (LPQs). Sometimes it is dubbed a 'miniblazar'
since it shows polarization flares similar to the true HPQs, but only
up to the 1\% level, and almost hidden by the non-polarized flux
variations (\cite{imp89,val90,val91}).

It is one of the best studied quasars and has been detected and
observed in every energy band from radio to \g -rays. It was first
detected as a very bright radio double source
(\cite{sch63}). Component A of the double source was identified with
the optical quasar (\cite{sch63,con82}) and has a very flat  
radio spectrum ($S \sim \nu^{\alpha _r}$) with a spectral index \ar\ =
-0.01 \+-\ 0.07 between 2.7 and 5 GHz (\cite{kue81}).  Component B is
associated with an optical jet and its radio spectral index is much
steeper: \ar\ $\approx$ -0.7. Its apparent magnitude in the optical is
$m_v$ = 12.5 mag. The optical jet is one-sided and faint.

\3c was first identified as an X-ray source by Bowyer \etal (1970).
Its luminosity in the 2 - 10 keV band is $\sim$\10{46} erg/s and the
X-ray spectral index in this energy band varies in the range -0.25
$\le$ $\alpha_x$ $\le$ -0.5 (\cite{cou87}). An X-ray feature
coinciding with one of the enhancements in the optical jet was
discovered by the EINSTEIN-HRI (\cite{har87}).  Two detections of a
high energy \g -ray source in the Virgo region were reported with the
COS-B satellite in 1976 and 1978.  This \g -ray source was identified
with \3c because of the positional coincidence (\cite{swa78,big81}).

During the all-sky survey of the Compton observatory in Phase~1 (1991
May to 1992 November), \3c was detected by all four instruments
aboard the observatory (BATSE:
\cite{pac94}; OSSE:  \cite{mcn94}; COMPTEL: \cite{her93,wil95}; EGRET: 
\cite{vM93}).
The main result from Phase~1 is the presence of another maximum in the
energy output in the 1-10 MeV range of the electromagnetic spectrum
which is about as high as the maximum of the UV-Bump (\cite{ggl94}, 1995).
 
Because \3c is such a well-studied, bright object from radio through
X-rays it seems to be well suited for detailed studies in order to
learn more about the physical processes taking place in quasars.
Intensive studies of the variability have already been performed in
order to look for correlations between different energy bands
(\cite{cou87,rob93}). These studies found correlations
between the infra-red (1.25\um (J); 1.65\um (H); 2.2\um (K); 3.8\um
(L$^\prime$); 800\um) and shortest radio wavelengths (1.1 mm and 3.3
mm). However, these correlations were not seen for every flare but
only in some (\cite{rob93}).

Since those studies did not include the high energy \g -ray range,
an international campaign was organized to
observe \3c simultaneously at all wavelengths from radio through MeV
and TeV \g -ray energies during Phase~3 (1993 August 17 to 1994
October 3) and Cycle~4 (1994 October 4 to 1995 October 3) of the
Compton Observatory mission, in the hope that it
would be possible to discriminate between the various models which
have been developed in order to explain the \g -ray emission from
blazars: (i) the synchrotron self-Compton (SSC) model (e.g. 
\cite{mar92,blo92},1993); (ii) the 
inverse Compton process on external photons (EC-models) which could be 
either photons from the accretion disk (\cite{der92}) or reprocessed photons
from the broad-line region (\cite{sik94,don95}), or
(iii) synchrotron emission from ultra-relativistic electrons and
positrons produced in a proton-induced cascade (\cite{man92,man93a}).
We are aware that there are even more models. But we can not consider
all these models here since it would go beyond the scope of this paper
which is mainly to present the data from the multiwavelength campaign.
Hence, we concentrate only on the three basic models. For an overview of
models see e.g. von Montigny \etal (1995) and references therein.

In this paper we describe observations of \3c with EGRET
during Phases~2, 3 and Cycle~4 of the Compton observatory mission
(\S2) as well as simultaneous or quasi-simultaneous observations
across the entire electromagnetic spectrum during Phase~3 (\S3). The
results are given in \S4. In \S5 we present a discussion of the results.

\section{EGRET observations and analysis}
 
A detailed description of the EGRET instrument is given by Kanbach et al. 
(1988). The instrument calibration, both before and after
launch, is presented by Thompson et al. (1993).
 
The analysis of the EGRET data (\cite{fic94}) used counts and exposure
maps for photon energies for different energy intervals as well as the diffuse
\g -ray background predicted by the standard EGRET analysis software from HI
and CO distributions (\cite{ber93,hun96}).
The maps containing all events with energies $\ge$ 100 MeV  were used for the
detection and determination of the position of the source in order to avoid 
the rather broad point spread function ($\ge$ \deg{5}) below 100 MeV. For the
determination of the spectrum of the source, the maps containing 10 standard
energy intervals were used.
 
This analysis used a maximum likelihood method which simultaneously
gives the best fit of the diffuse background to the data (\cite{mat96}). 
Prominent EGRET sources other than \3c in the viewing period under
consideration were added iteratively to the diffuse background model.
Throughout the analysis a photon spectrum power law index of \Gg =2.0 was
assumed for the spectra of the sources which is a typical spectral index for
the strong EGRET blazars (\cite{vM95,chi95}). 
The formal significance of a source detection in standard deviations is
determined from the square root of the likelihood test statistic
\ts (\cite{ead71,mat96}) which is given by
two times the natural logarithm of the ratio of the maximum likelihood
values for the alternative and the null-hypothesis.

\subsection{Time variability} 

Table 1 lists the observations, the fluxes and significances from the EGRET
observations in phases 1, 2, 3 and cycle 4.  When the source has \ts $< 9$
(corresponding to a formal significance of $<$ 3\s ) it is regarded as not
detected and 2\s\ upper limits are given.

After the initial detection of \3c by EGRET in 1991 June (viewing period
(VP)~3) in phase~1 (\cite{vM93}) its flux (always for energies $\ge$ 100
MeV) had decreased in strength during later observations in 1991 October
(VP 11) and in 1992 December through 1993 January (VP 204 through 206) 
in phase~1 and~2 (\cite{vM93,sre96}); only upper limits could be derived (see
also Table~1 and Fig.~\ref{fig:time_hist}, bottom panel).  
Then in phase~3 during 1993
October to 1993 December (VP's 304 - 308.6) and 1993 mid-December to
1994 January (VP's 311 - 313) the flux of \3c increased from about
2.1\c10{-7}\pcm2sec\ to approximately 5.6\c10{-7}\pcm2sec\ between
1993 October 19 and 1993 December 1 (VP's 304 through 308.6). This is
an increase by about a factor of 3 within 43 days (Fig.~\ref{fig:time_hist}, 
bottom panel) and it is the only time so far that EGRET has seen \3c to be at
about the same flux level (6.0\c10{-7}\pcm2sec ) as during the COS-B
observations (\cite{swa78,her81}).  Fourteen days later, and again in
Cycle~4 (1994 November 21 through 1995 January 3) it returned to a
quiet state.

\subsection{Spectrum}

We derived spectra from each viewing period and sums of viewing periods
in which \3c was detected with a significance greater than 3\s . 
In order to determine the spectra the estimated number of source counts in ten
observed energy intervals was determined by a likelihood analysis 
(\cite{mat96}).
  
The simplest
spectral model that adequately fit the data was a power law of the form
$$ dN/dE  = N_o(E/E_o)^{-\Gamma_\gamma} $$ 
where $E_o$ is the energy scale
factor chosen so that the statistical errors in the power law
index, \Gg, and the overall normalization, $N_o$, are uncorrelated.  

The values of the parameters $N_o$, \Gg\ and $E_o$ are given in Table
2 for the different observations. The gamma-ray photon spectral index
was found to vary from viewing period to viewing period from \Gg\ =
(3.20\+-0.54) to \Gg\ = (2.20\+-0.22).  The errors are determined from
$\Delta$\chisq = 1 (see Dingus \etal 1996).

Figure~\ref{fig:spectra} compares the spectra from \3c during VP 305 and 
VP308.6 with
each other. These are the observations where \3c had the lowest and the
highest flux in phase 3, respectively.  It can be seen that the
increase in flux results from the hardening and pivoting of the
spectrum around the low energy end.
 
There appears to be a correlation between the spectral index and the
integral flux above 100 MeV in phase 3 (Fig.~\ref{fig:spec_comp}). The
linear correlation coefficient between these two variables is -0.91,
and the significance level at which the null hypothesis of zero
correlation is disproved is 1.1\%.  However, if one includes the data
from VP 3 and COS-B, the linear correlation coefficient changes to
about -0.72, corresponding to a probability that this set of data was
drawn from a uniform distribution of about 4.3\%.  The significance of
the correlation is not very high since the errors on both variables
are rather large. Since this correlation is not yet compelling we will
use the average EGRET spectrum derived from the sum of the viewing
periods 304 through 308.6 for the rest of the paper.  Nevertheless,
there is evidence that the \g -ray spectra harden when the source flux
increases. There are indications for this behaviour not only from \3c
but also from other \g -ray sources. M\"ucke \etal (1996a) did a
statistical analysis of this relation with all EGRET sources for which
spectra and fluxes were available (e.g.  0528+134, \cite{muk96};
1222+216, \cite{sre96}; 3C~279, \cite{kni93}). They also find that the
average source appears to have a harder spectrum at high \g -ray
states. The chance probability is of the order of \10{-5}.

\section{Multiwavelength observations}
Soon after the discovery of the June 1991 flare in 3C~279 and the realization
that most of the \g -ray blazars are highly time variable, it was recognized
that simultaneous observations of these sources across the entire
electromagnetic spectrum are of crucial importance for the understanding of
their emissions.  This led to an international campaign to observe \3c
simultaneously at as many wavelengths as possible in 1993 and 1994. The
following section reports on the results of these observations.

\subsection{Radio}
Reich et al. (1993) have already described the multifrequency observing
method used at the Effelsberg 100-m telescope to monitor variable
sources as detected by EGRET. The observations result in
quasi-simultaneous flux density measurements at 2.7 GHz, 5 GHz and 10.7 GHz.

Observations at 22 and 37 GHz were made at the Mets\"ahovi Radio
Research Station. The observing procedures are described in detail in
Ter\"asranta \etal (1992). The data presented here are weekly mean
values.
 
Observations were also made with the University of Michigan's 26-meter
telescope at 4.8, 8.0 and 14.5 GHz. A description of the
data reduction is given in Aller \etal (1985). The fluxes are daily
averages. Table 3(a-f) lists the radio data used here.

\subsection{Millimeter and sub-millimeter}
The 3 mm and 1.3 mm (90 GHz and 230 GHz, respectively) observations
were made with the Swedish-ESO Submillimetre Telescope (SEST) at the
European Southern Observatory site of Cerro La Silla, Chile.

For the 3 mm observations  a dual polarisation Schottky receiver in a
double sideband mode was used. 
For the 1.3 mm observations a Schottky receiver and a wide band
acousto-optic spectrometer were used initially, but later a single
channel bolometer was used for most of the sessions.
The bolometer had a bandwidth of about 50 GHz, centered at 236 GHz.
To convert the intensities into flux densities, the measurements were
calibrated against planets, with 3C~274 as a secondary calibrator 
(\cite{tor96}).

Data were also obtained 
at wavelengths of 450, 800, 1100, 1300 and 2000 microns  using the JCMT
at Mauna Kea (\cite{dun90}).
Tables 4(a-f) summarize the results from the observations in the  
mm- and sub-mm wavebands. 

\subsection{IR and optical}

Observations have been carried out in UBVRIJHK with sparse sampling at
the Swiss 0.7m telescope on La Silla, the 0.7m telescope in Heidelberg
and with UKIRT in Hawaii using standard CCD detectors and a NICMOS 3
array camera. The data are shown in Table 5. Observational procedures
and data calibration has been carried out as described in Courvoisier
\etal (1987).

\subsection{UV/EUV}

\3c\ is observable by IUE only during the time intervals mid-December
-- mid-February and May -- mid-June.   There were observations of
\3c\ with IUE on 1993 January 8 (2 SWP respectively 3 LWP spectra coadded) and
1994 June 20 (4 SWP, 4 LWP) as well as from 1995 January 3 to 12
(20 SWP, 21 LWP). The average exposure for SWP spectra was $\sim$ 30
minutes and for LWP spectra $\sim$ 25 minutes.

We have also obtained EUVE coverage of 3C 273 with the EUVE Deep
Survey/Spec\-tro\-meter during the 1994 January 8 to 14 (modified
julian dates: 49360 - 49366) time frame for a total of 205,219 s
(effective exposure time: 130,093 s). The total photometric flux in
the range 67 -- 178
\AA\ was (7.2\+- 0.081)\c10{-2} counts/s. For details about the analysis 
see Ramos \etal (1996).
   
Table 6 lists the results of these observations.  A reddening correction for
interstellar absorption of E(B-V) = 0.03 has been applied to the
values from
the reddening law of Seaton (1979). The errors in fluxes are obtained by also
taking into account the calibration error of IUE ($\sim$ 5-10\%).  

\subsection{X-rays}

ASCA observed \3c for calibration purposes on five different occasions
namely on 1993 December 16, 19, 20, 23 and 27.
Assuming a single power law with the Galactic absorption ($N_H$
= 1.8\c10{20} H-atoms/cm$^2$) in the 0.5-10 keV range a
spectrum for \3c could be derived from these measurements. Details of the 
ASCA instrument and its calibration are given in   
Makishima \etal (1996) and references therein.
Table 7 contains the results which were obtained using the 1994
versions of the instrumental response and efficiency functions for the
analysis.

\subsection{High energy observations}

The OSSE instrument is sensitive to \g -rays in the 0.05 -- 10 MeV range. It
consists of four identical but independently positionable detector systems
providing an orientation range of \deg{192}. For details about the
instrument and analysis procedures, see e.g. Johnson \etal (1993, 1995).
OSSE observed \3c simultaneously with COMPTEL and EGRET only for 7
days from 1993 December 20 to 27 (VP 312). The results
from these observations are given in Table 8.

The COMPTEL energy range (0.75 -- 30 MeV) overlaps with the OSSE
energy range. Detailed descriptions of the instrument can be found in 
Sch\"onfelder \etal (1993).
COMPTEL has also detected \3c and Table 9 gives the fluxes derived by a 
maximum likelihood method (for its application to COMPTEL data see 
\cite{dBo92}). The spectrum obtained is the average over two distinct time
periods: 1993 October 19 to December 1 and 1993 December 13 to 1994
January 3 (VP's 304 -- 313).

The Whipple Telescope observed \3c during several epochs. The
observation  closest in time to the EGRET observations is from 1994 March.
\3c was not detected during that observation. Therefore, only an upper limit
could be provided. For energies $\ge$ 0.3 TeV the flux limit was $<$
1.7\c10{-11}\pcm2sec .  Assuming
an $E^{-2}$ power law this flux value corresponds to a flux density
\Fn\ $<$ 1.1\c10{-2}pJy.

\section{Results}\nobreak
\subsection{Multifrequency Variability}

In the case of \3c it seems that the gamma radiation is possibly
related to the slower radio outbursts at 14.5, 22 and 37 GHz. There
are two slower outbursts at 22 and 37 GHz, the first peaking around
1991.67 (MJD $\sim$ 48470) and the second around 1993.96 (MJD $\sim$
49330).  The second outburst peaks first at 37 GHz then at 22 GHz and
finally at 14.5 GHz. The second outburst does not look as impressive
because it is buried within the decline of the very strong first
outburst; nevertheless it still represents a major radio burst.  All
the detections of \3c with EGRET occured during the rising parts or at
the maxima of the major, slower radio outbursts. Most of the
non-detections were during the declining part of these burst
components or during a very early stage of the outburst
(Fig.~\ref{fig:time_hist}, 2nd and 4th panels).

The \g -emission disappeared during 1993 December.  This seems to
occur just after the second radio burst component is peaking. According
to Valtaoja \& Ter\"asranta (1995) EGRET detections occur in general
during the rising part of the 22 and 37 GHz radio flare, and once the
radio reaches its peak, the gamma radiation ceases. 

While it is fair to say that the gamma-ray-bright
phase coincides temporally with radio activity, there is no way to
prove or disprove the hypothesis of `related' activity on
the basis of the data shown here. Statistically
there appears to be good evidence that gamma-ray detections occur during
radio outburst rises, but with the undersampling of the gamma-ray data
one cannot make claims for correlated gamma-radio activity. The entire
period discussed here corresponds to a general decline at cm-wavelengths
on which some more rapid fluctuations are superposed.  
Also, note that while 3C 273 is very bright at radio wavelengths, in the
gamma-ray region it has been only moderately strong during detections; there
is no clear correlation between gamma-ray flux and radio-flux in
general.

It is interesting to note that the intensity of \3c appears to have
gone through a local, very flat minimum for wavelengths $\le$ 1.3mm
(Fig.~\ref{fig:time_hist}, 3rd panel) during the observed increase at
14.5, 22 and 37 GHz and in \g -rays. A $\chi^2$ test shows that the
spread in the sub-mm fluxes in the time range MJD 49300 -- 49400 is
consistent with constant fluxes even for the 1.1 mm data
(Fig.~\ref{fig:zoom}). However, the period just before and during the
first part of the gamma outburst was not observed because of sun
constraints.

In the IR/optical no significant variations have been recorded within
the poor sampling throughout the entire period. The range of
variations found by Courvoisier
\etal (1987) in the optical/IR range was larger, but the average values
are compatible. Likewise, the near-IR and optical continuum slopes
(as measured by nearly simultaneous JHK, UBV or BVRI sequences) remained
constant throughout the period and comparable to the average slopes.
During several epochs with non-photometric conditions variations on
short time-scales were searched for by differential photometry
(comparing fluxes to constant stars within the same frames). These
data had a sampling of about 1 hour$^{-1}$. They are not flux
calibrated and not listed in Table 5. No indications for rapid
variations in V and R were found to a level of 1.2 \%.
 
\subsection{Multifrequency Spectrum}

Unfortunately, not all of the multiwavelength observations were 
truly simultaneous with the EGRET observations between 1993 October 19
and 1994 January 3 (MJD 49279 - 49355). For example, the closest IUE
data available are from 1994 January 8 (MJD 49360; see Table 6) and
the OSSE spectrum is from 1993 December 20 to 27 (MJD 49341 -- 49348;
Table 7). Anyway, one should
bear in mind that the time spans involved in deriving a spectrum from
a source are very different from one wavelength range to the other. In
the \g -ray range for example the spectrum has been derived from data
accumulated over at least one week, while in the other wavelength
ranges spectra can be derived on timescales of one day and less. For
this reason alone the \g -ray spectrum can never be truly simultaneous
to the other multiwavelength spectra. In order to be truly
simultaneous the sampling times should be of the same order.

The overall spectrum of \3c during the phase 3 observations
(Fig.~\ref{fig:mw_spectrum}) is similar to that during 1991 June (VP
3) in phase 1 (\cite{ggl95}).  The multiwavelength energy density
spectrum (\nFn\ spectrum) shows probably four maxima: the first around
3\c10{11} Hz (corresponding to the mm --- sub-mm range), the second
maximum (although not observed) must be between \10{12} Hz and
$\approx$\10{14} Hz, the third maximum is the ``blue bump'' at about
3\c10{15} Hz and the fourth maximum is reached in the MeV region
between 1 and 10 MeV. It is the first maximum, in the mm -- sub-mm
range, which was at a relatively low level during the \g -ray
`flare'. So far, it is not at all clear whether this observation fits
the theoretical explanations for the production of the high energy \g
-rays.

In order to determine the break energy and the change in spectral
index between the hard X-rays and the high energy \g -rays, the energy
spectrum beyond 2 keV has been fitted with the same broken power law
already used by Lichti \etal (1995). The measured energy spectrum is rather 
well represented by the following (empirical) function (Fig.~\ref{fig:break}; 
the reduced 
$\chi^2$ for this fit is $\chi^2/dof$ = 1.04 with 15 degrees of freedom):
$$F_{\nu}=(1.66 \pm 0.87)\cdot 10^{-4} {{(E/E_b)_{MeV}^{-(0.629 \pm 0.038)}}
                        \over{1 + (E/E_b)_{MeV}^{(0.97 \pm 0.07)}}}\ [mJy]$$
Where $E_b$ has a value of (2.36 \+- 1.28) MeV. The fit function
indicates a steepening of the energy spectrum by 0.97 \+- 0.07  
for E $>>$ 2 MeV. This is somewhat more than the value from
Lichti \etal (1995) which was 0.8 \+- 0.03 for E $>>$ 1 MeV 
(Fig.~\ref{fig:break}).

\subsubsection{Model fits} 

In order to test the different models for the generation of \g -rays
in the jets of AGN we have fitted the observed spectrum with (i) a
synchrotron self-Compton (SSC) model, (ii) an external Compton (EC)
model, and (iii) a proton-initiated cascade (PIC) model. The basic
assumptions for the jet in all these models are the same as in
Blandford \& K\"onigl (1979): a conical relativistic and magnetized
jet in which accelerated electrons produce a flat radio synchrotron
spectrum (owing to a synchrotron-self-absorption turnover frequency
decreasing inversely proportional to the jet radius) breaking by one
power in the mm-to-infrared range due to energy losses steepening the
accelerated electron spectrum.

(i) The SSC model: The relativistic jet SSC calculations are described
by Marscher and Travis (1996).  The jet is modeled as a truncated cone
with a power-law electron spectrum injected at the inner radius. The
density falls off as $1/{\rm r}^2$, the magnetic field as 1/r, and the
electron energies decay from adiabatic, synchrotron, and inverse
Compton losses. This provides a smooth-jet approximation to
the knotty structure observed with VLBI.

For the model fit, the following parameters have been fixed: Opening
half-angle of the jet (\deg{0.5}), angle between the jet axis and the
line of sight (\deg{6}), bulk Lorentz factor 9.3, minimum injected
electron energy corresponding to \g$_{min}$ = 100.  The following
parameters were determined by the fitting procedure: Injected
power-law of electron energy distribution s = 2.3, ratio of randomly
oriented to axial component of magnetic field: 1.5; the values at the
injection point of the parameters that change with radius are:
Cross-sectional radius r = 0.055 pc, magnetic field B = 0.023 G,
density of relativistic electrons n$_{\rm e}$ = 180 cm$^{-3}$, and the
maximum injected electron energy \g$_{max}$ = 2.5\c10{4}.

Hence, 6 parameters were varied until a good fit
(Fig.~\ref{fig:SSC_fit}) was obtained. We required that the model fit
the self-absorption turnover frequency and the overall spectral
shape. This places strong constraints on the values of the critical
parameters. Nevertheless, we cannot be certain that the fit is the
best one possible. The $\chi^2$ for this fit is 270 with 22 degrees of
freedom. This applies to the frequency range 3.7\c10{10} to 2\c10{14}
Hz and 1\c10{18} to 1\c10{24} Hz.  There was no attempt to fit the
optical-UV spectrum, since it is obviously dominated by the big blue
bump. The main contributions to the high $\chi^2$ are several points
that are poorly fit but have very low observational uncertainty. The
$\chi^2$ test indicates that the fit is not particularly good in
detail.  However, we are only using an ideal model with a small number
of free parameters that can provide a global fit rather than a more
complex model designed to fit each point as closely as possible.

We therefore conclude that an SSC-emitting relativistic jet model
can reproduce the general shape of the multiwaveband spectrum of 3C 273.
This is true even though the physics of the energy losses and
radiative transfer are fixed ab initio.

(ii) The EC-model: In this model, the (steady) electron distribution
$N$(\g ) has been derived using the continuity equation as well as
considering cooling, escape, pair production etc. (\cite{ghi96}). In
this case one obtains a broken power law because electrons below the
minimum injected \g$_{min}$ have a \g$^{-2}$ distribution, while
electrons above \g$_{min}$ have \g$^{(-s-1)}$, where $s$ is the slope
of the injected electrons. In this case it is assumed that electrons
are injected continuously, with a power law energy distribution with
slope $s=3.3$ between \g$_{min}=80$ and \g$_{max}=10^4$.  This results
in an equilibrium distribution which is a broken power law, with index
2 for \g $<$ \g$_{min}$, and around 4.3 above.  The blob sees photons
coming from outside, [i.e. photons from the broad line region (BLR)],
distributed as a diluted blackbody.

Figure~\ref{fig:EC_fit} shows the result of the fit of this model to
the data.  The parameters of the model are: Dimension of the source
$R$ = 2\c10{16} cm, magnetic field $B=5.9$ G, injected luminosity
$L_i$ = 3.7\c10{44}erg/s (intrinsic), beaming factor $D=6.5$ (assumed
to be equal to the bulk Lorentz factor) and the ratio of external
radiation energy density to magnetic energy density $U_{ext}'/U_B$ =
11/5.5 = 2 (in the comoving frame of the blob). The blackbody in the
figure corresponds to the relevant disk emission, which illuminates
the BLR.  The $\chi^2$ for this fit [considering only the points above
log($\nu$)=15, and excluding the point in the soft X-rays at
log($\nu$)=16.477], is 108.84. Assuming 6 interesting, adjustable
parameters ($R$, $B$, \g$_{min}$, \g$_{max}$, slope of the injected
electron distribution, and $R_{BLR}$), the reduced $\chi^2$ for this
fit is $\chi^2/dof= 7.78$.

The Compton spectrum is not completely smooth, because it is the sum
of internal SSC and external Compton.  One can see the contribution of
the internal SSC at \n $\sim$ \10{16} Hz.  Electrons with \g $<$ 100
emit self--absorbed synchrotron radiation, and therefore one does not
see the synchrotron emission of the electrons below the break.  

Due to the simplicity of this one--zone, homogeneous model, we made no
attempt to fit the radio, the far IR and the soft X--ray excess
emission.  Additional, larger, non-thermal components are needed to
model the far IR and the radio data, and another (maybe thermal)
component must be responsible for the soft--ray excess.  Furthermore,
there is no intention to model the UV--bump emission correctly with
some theory.

The value of the derived $\chi^2$ can be considered only a very rough 
measure of the goodness of the fit, given the simplicity of the model, 
the maybe unrealistically small error bars on some data points (especially 
in the MeV band) and the fact that not all data points are strictly
simultaneous.

(iii) PIC-model: Figure~\ref{fig:PIC_fit} shows a fit to the \3c
multifrequency data adopting the proton blazar model (\cite{man93b}).
In this model, $\gamma$-rays emerge as the end-product of cascades
initiated by ultra-relativistic protons suffering inelastic collisions
with low energy synchrotron photons, which appear in the proton rest
frame with energies above threshold for secondary pair and pion
production. The proton-initiated emission is treated in a one-zone
approximation in which the cascades are assumed to occur only at the
jet radius where the infrared synchrotron photons (acting as a target
for the ultra-relativistic protons) become optically thin.  In the
present model fit, the jet radius computed from the fit parameters is
$r_\perp = 10^{17}$~cm (the distance to the central black hole is
undetermined).

Fitting the low energy part of the spectrum as electron synchrotron
emission yields a jet Lorentz factor $\gamma_{\rm j} = 8$, angle to
the line-of-sight $\theta = 7^\circ$, opening angle of the jet $\Phi'
= 2^\circ$ and a relativistic particle luminosity $L_{\rm j} =
1.4\times 10^{46}$~ergs~s$^{-1}$.  The proton-to-electron ratio was
forced to be $\eta = 100$ allowing for a calculation of the
equipartition magnetic field strength $B_{\rm eq}' = 0.8$~G and the
synchrotron break frequency in the comoving frame $\nu_{\rm b}' =
2\times 10^{10}$~Hz.

Fitting the high-energy part of the spectrum yields the ratio of the
proton and electron cooling rates ($\xi= 0.15$) considering that the
cascade luminosity emerging in the X-ray and $\gamma$-ray bands equals
$L_{\rm p} = \eta\xi L_{\rm e}$ where $L_{\rm e}$ denotes the primary
electron synchrotron luminosity (radio-to-UV).  From the value of
$\xi$, one obtains a proton maximum Lorentz factor of $\gamma_{\rm
p,max} = 3\times 10^{10}$.

In addition to the jet spectrum, an accretion disk spectrum with
$S_\nu\propto \nu^{+1/3}$ in the optical-to-UV range has been assumed.
The inferred thermal-to-nonthermal luminosity ratio is of order unity.
The disk spectrum turns over steeply at $20$~eV turning into a power
law $S_\nu\propto \nu^{-2}$.  This kind of soft X-ray spectrum is
expected to emerge from a disk covered by a marginally optically thick
jet base (\cite{man95}).

Also $\gamma$-ray attentuation by interaction of the jet $\gamma$-rays with
photons from the jet environment has been taken into account. This is
important, since the proton blazar spectrum tends to overproduce $\gamma$-rays
above $\sim 100$~MeV, and especially at TeV energies.  
In the fit, attenuation by
diffuse intergalactic light and by scattered disk radiation has been taken into
account.  The former leads to a quasi-exponential cutoff at $\sim 7\times
10^{11}$~eV (\cite{man96}), the latter is characterized by an optical
depth $\tau_{\gamma\gamma} = 0.01(\epsilon/{\rm 100\thinspace
MeV})^{1/2}L_{46}\tau_{-2}$  (\cite{der94}) leading to a
steepening by $\Delta s = 0.5$ in the GeV range ($L_{46} = 3$,
$\tau_{\rm -2} = 5$).
Thus, although in the proton blazar model $\gamma$-rays could in
principle be produced at an arbitrary distance from the central black
hole (provided that the jet is collimated enough to produce infrared
synchrotron photons of high density), the spectrum of \3c indicates
that the flux of $\gamma$-rays is attenuated by traversing the central
radiation field. The occasionally observed flattening of the
$\gamma$-ray spectrum is naturally explained as the signature of a
$\gamma$-ray emitting shock traveling along the jet away from the
central source of thermal radiation, thereby experiencing a decreasing
external pair creation optical depth.

Owing to the coarse construction of the model, fine details of the
spectrum in the radio-to-infrared and optical-to-UV bands are not well
reproduced.  They could be fitted with much higher accuracy taking
into account jet inhomogeneities and an accretion disk for which
published models exist.  Such refinements for the proton-initiated
cascade part of the model would probably also remove the significant
residual at 1-10~MeV which is responsible for the moderate
$\chi^2/\nu=32/20$ of the fit in the X-ray-to-$\gamma$-ray regime.

\section{Discussion}

The increase in \g -ray flux is much slower for \3c than that observed
from 3C~279 during its 1991 June flare where the flux increased by a
factor of 4 within about 7 days (\cite{kni93}).  On the other
hand the difference in peak flux from the two sources is about a factor
of 7. While in 3C~279 we may have seen only the top of the flare, it is
more likely that in the case of \3c we have seen only the initial rise
of a flare. One could therefore argue that a \g -ray flare might
consist of two parts: a slowly increasing part (as observed in \3c)
followed by a more eruptive part with a steep decrease at the end as
observed in 3C~279 (\cite{kni93}). Such an eruptive part could
have easily fit into the two weeks following 1993 December 1 when \3c
was not observed by EGRET.

But it could also be that the shape and intensity of the flare is
related to the temporal behaviour at lower frequencies: In the case of
3C~279 there was a possible correlation of the \g- ray flare with a
short synchroton flare in the R-band (\cite{mch96,har96}) while in the 
case of \3c the increase in flux might have
been correlated with the much slower second radio
outburst components at 22 and 37 GHz.  A similar behavior was observed
for the quasar PKS 0528+134 where also a radio outburst at 22 and
37 GHz followed the \g -ray flare (\cite{poh95,muk96}). Further flares 
have to be observed before any of these
hypotheses can be confirmed.

As already noted by Courvoisier \etal (1987, 1990) the temporal
behaviour is very complex in all wavebands with a few correlations,
none conclusive mainly because the light curves are undersampled
in most cases (especially at shorter wavelengths). Our observations
are no exception.  From the temporal behavior of \3c during this
observational campaign across the entire electromagnetic spectrum one
cannot deduce any correlation or anti-correlation of the high energy
end with the lower energies, except maybe for the 14.5 GHz, 22 and 37
GHz regime. Statistical investigations of the temporal behavior of
{\it all} the \g -ray emitting AGN seem to support a relation with the
22 and 37 GHz regime (\cite{val95,mue96b}) although there is no
one-to-one correlation.

The decrease of the flux in the sub-mm and mm-regime during the flux
increase in the \g -ray regime is probably coincidential, but still
represents a problem for the homogeneous SSC models in which changes
in intensity (though different in amount) should go in the same
direction for all frequencies. 

Courvoisier \etal (1987, 1990) already excluded the homogeneous SSC
models for \3c  because of the lack of mm-X-ray correlations at zero 
lag\footnote{Another very prominent quasar for which such a lack of
mm-X-ray correlation is known is 3C~345 (\cite{bre86})}. They
concluded that this lack of correlation could only be explained with
either two different electron components (one for the synchrotron, one
for the X-rays) {\it OR} a second source of photons which dominates the
photon energy density in the synchrotron region. This second source of photons
would naturally be explained by the EC-models, the second electron
component would naturally follow from the proton-induced
cascade models where the X- and \g -rays can be produced by completely
different particle populations. 

Although the increase of the \g -ray intensity was larger than the increase in
the synchrotron emission (except for the decrease in the sub-mm and mm-regime),
which is qualitatively consistent with the expectations from the SSC models
(e.g. \cite{mar94}), the SSC model has to assume either inhomogeneous emission
regions or a broken power law for the spectrum of the relativistic electrons
(\cite{ghi96}) in order to account for the -- in general -- two peaks
(one in the IR-optical and one in the \g -ray regime (see e.g. \cite{vM95})) 
in the overall spectral energy distribution of the \g -ray emitting AGN.
\3c has at least three (maybe even four) peaks in its  spectral energy
distribution, indicating an even more complicated situation and maybe even as
many as four different emission mechanisms (\cite{cou87}). 

The attempts of fitting the observed multiwavelength spectrum of \3c with
theoretical models also show that the spectrum can not be well represented
without the contribution of several components. For example, the ``big blue
bump'' peaking at $\approx$3\c10{15}Hz may be indicating the presence of a
massive accretion disk providing copious possible target photons for the
relativistic particles in the jet.  All three models are capable
of representing the MeV-Bump within the errors. In the
X-ray/\g -ray range, the data cannot be fit well by a smooth SSC
spectrum within the 1 \s\ error bars but including the big blue bump
and first order Comptonization of these photons might improve the
situation in that regime. The EC-model has some difficulties to
account for the low energy range between \10{9} $<$ \n $<$ \10{14} and
also the extreme UV range without assuming at least additional
emission sites.  The PIC-model yields the best overall fit of the
multiwavelength spectrum over more than 17 decades with only three
necessary components.  Another possibility to generate the high
\g -ray emission which has not been mentioned yet is the
Comptonization of the ``big blue bump'' by a monoenergetic
relativistic electron outflow involving multiple inverse Compton
scattering (\cite{tit95,ram96}).

Although this campaign has the by far best coverage across the entire
electromagnetic spectrum the data are still insufficient to
discriminate between all the possible emission models for the high \g
-ray emission. What is still missing is a truly {\it simultaneous} and
{\it regular} monitoring of \3c (and other blazars) at all frequencies
before, during and after a flare in high energy \g -rays. The more we
learn about \3c and other \g -ray emitting quasars the more it is
becoming evident that \3c is far from being a standard quasar.

\acknowledgements

We thank John P. Travis for providing his code for fitting the
SSC model to the data. We also thank the referee John Mattox
for his valuable comments. 
The EGRET Team gratefully acknowledges support from the following:
Bundesministerium f\"ur Bildung, Wissenschaft, Forschung und Technologie
(BMBF), Grant 50 QV 9095 (MPE); NASA Cooperative Agreement NCC 5-93 (HSC); 
NASA Cooperative Agreement NCC 5-95 (SU).
This work was also supported by the Deutsche Forschungsgemeinschaft
(Sonderforschungsbereich 328). UMRAO is supported through NSF grant
AST-9421979. A.P.M. gratefully acknowledges partial support of the
work on this project at Boston University by NASA through grant
NAG5-2508.

\clearpage
\vsize=29.0 truecm

\figcaption[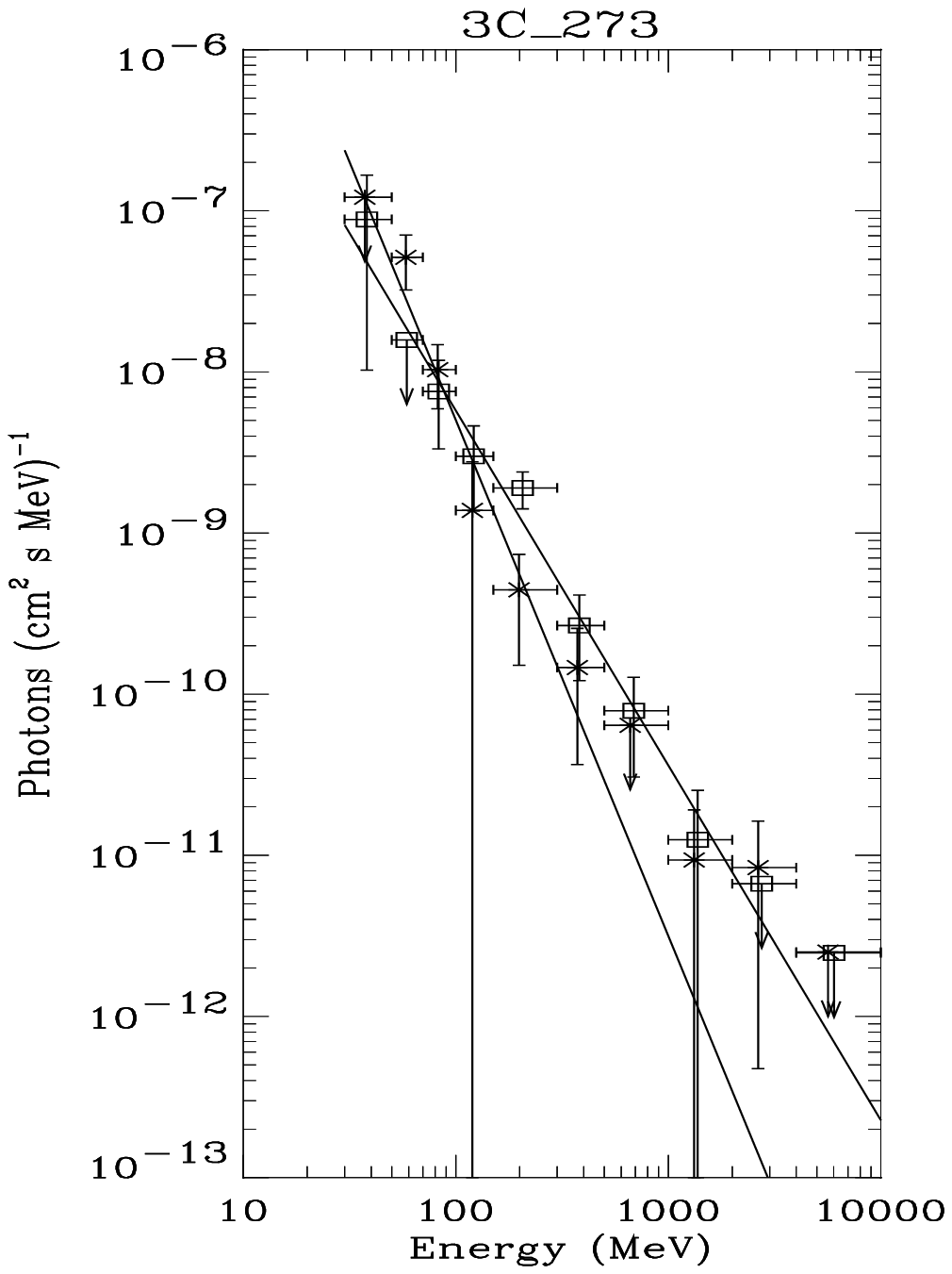]{Comparison of spectra
of \3c\ during VP 305.0 (asterisks) and VP 308.6 (squares). It can be
seen that the increase in flux comes from a hardening of the spectrum. 
\label{fig:spectra} }

\figcaption[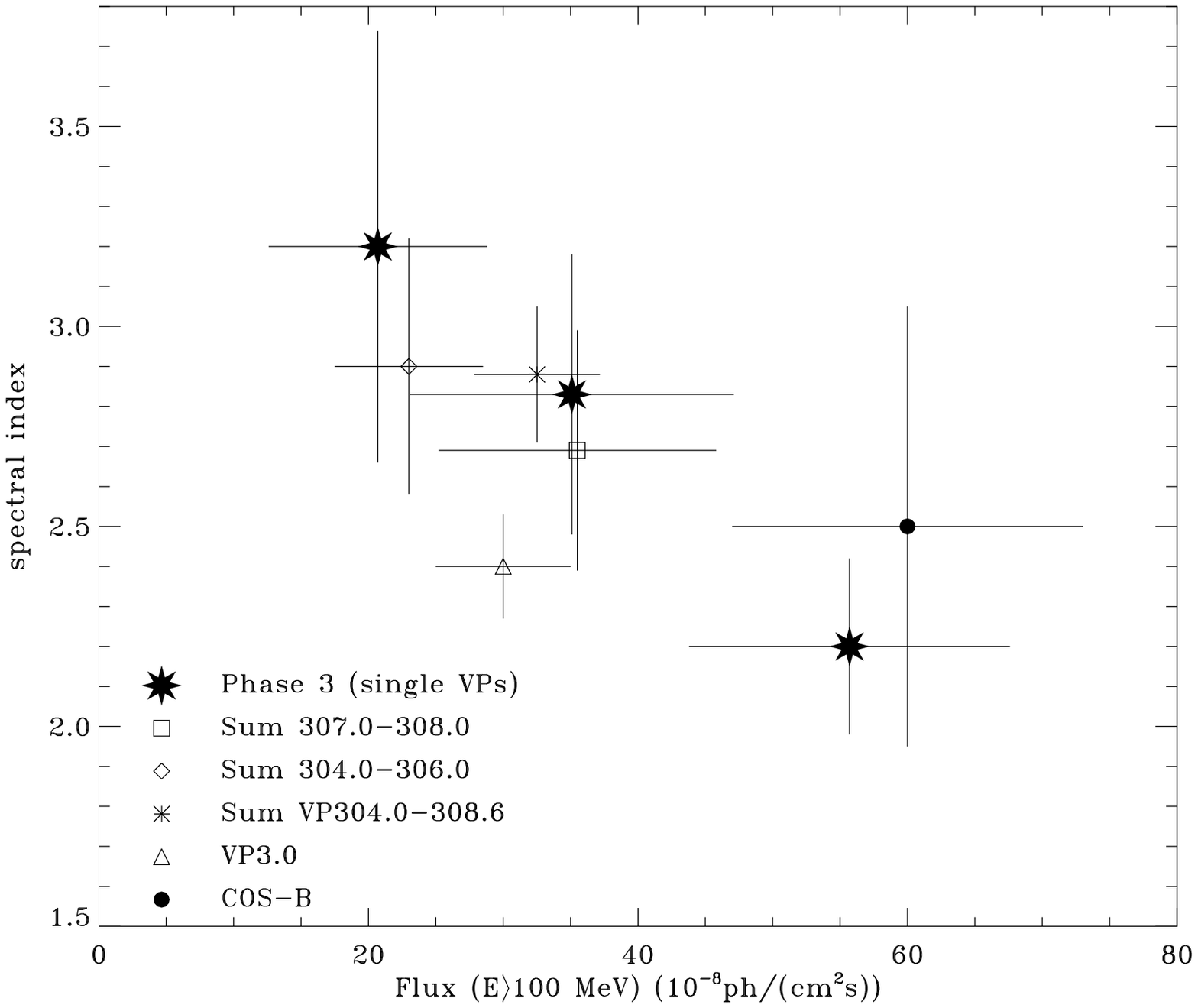] {Spectral indices
versus integral flux above 100 MeV for 3C~273. There is evidence that
the \g -ray spectra harden with increasing flux. \label{fig:spec_comp}}

\figcaption[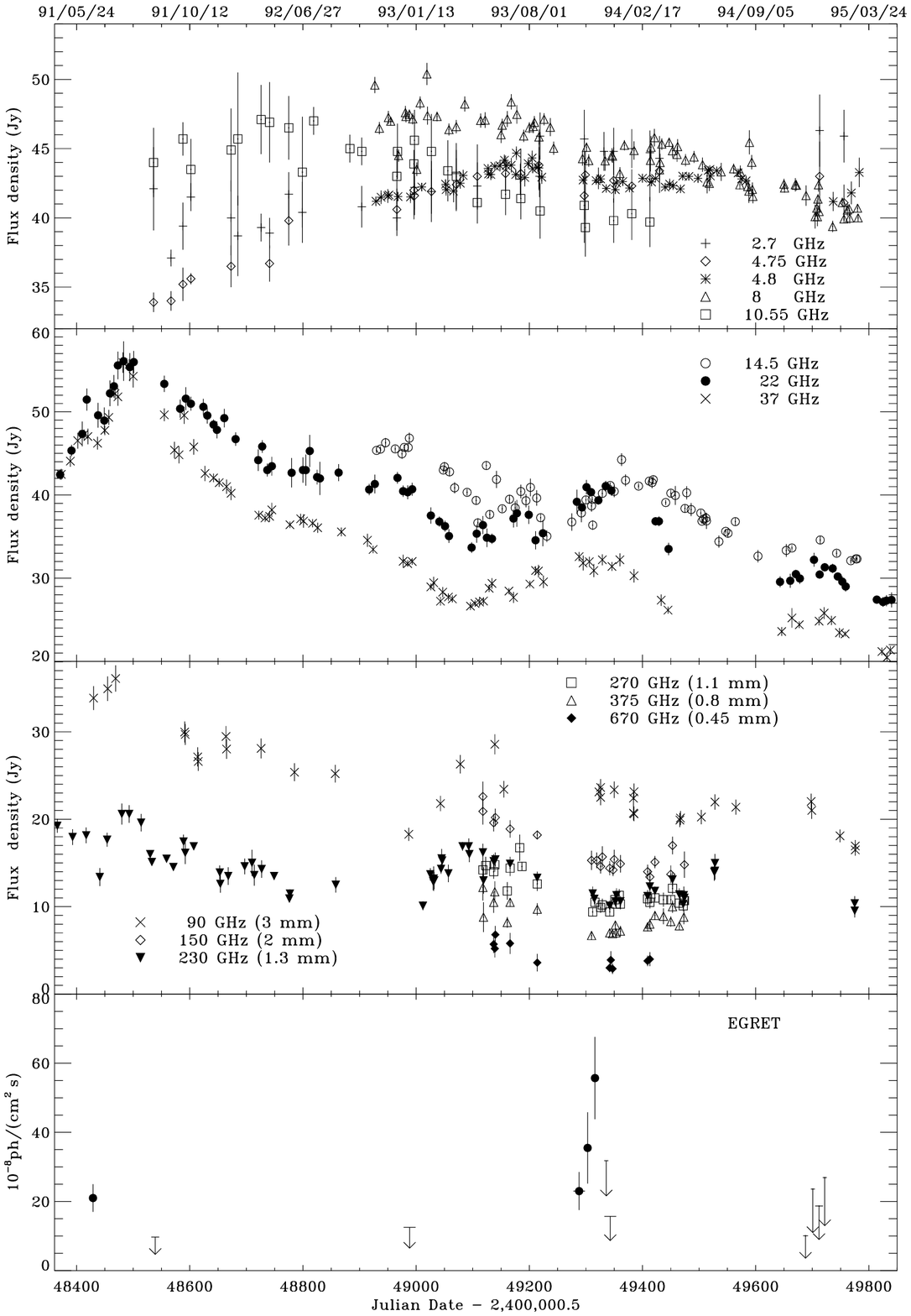]{Time histories of \3c. \label{fig:time_hist}}
 
\figcaption[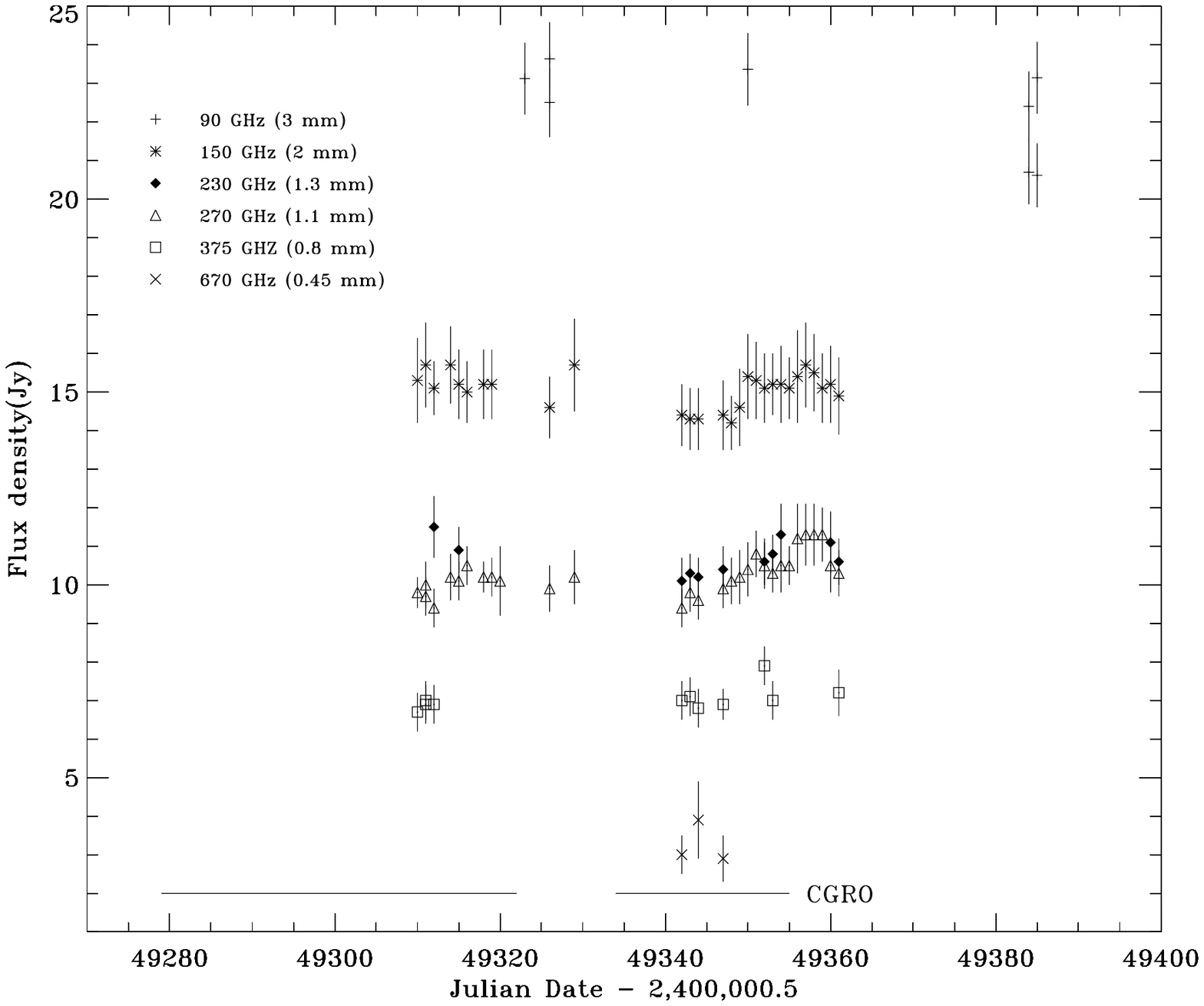]{mm- and sub-mm fluxes of \3c\
during the EGRET observations. \label{fig:zoom} }

\figcaption[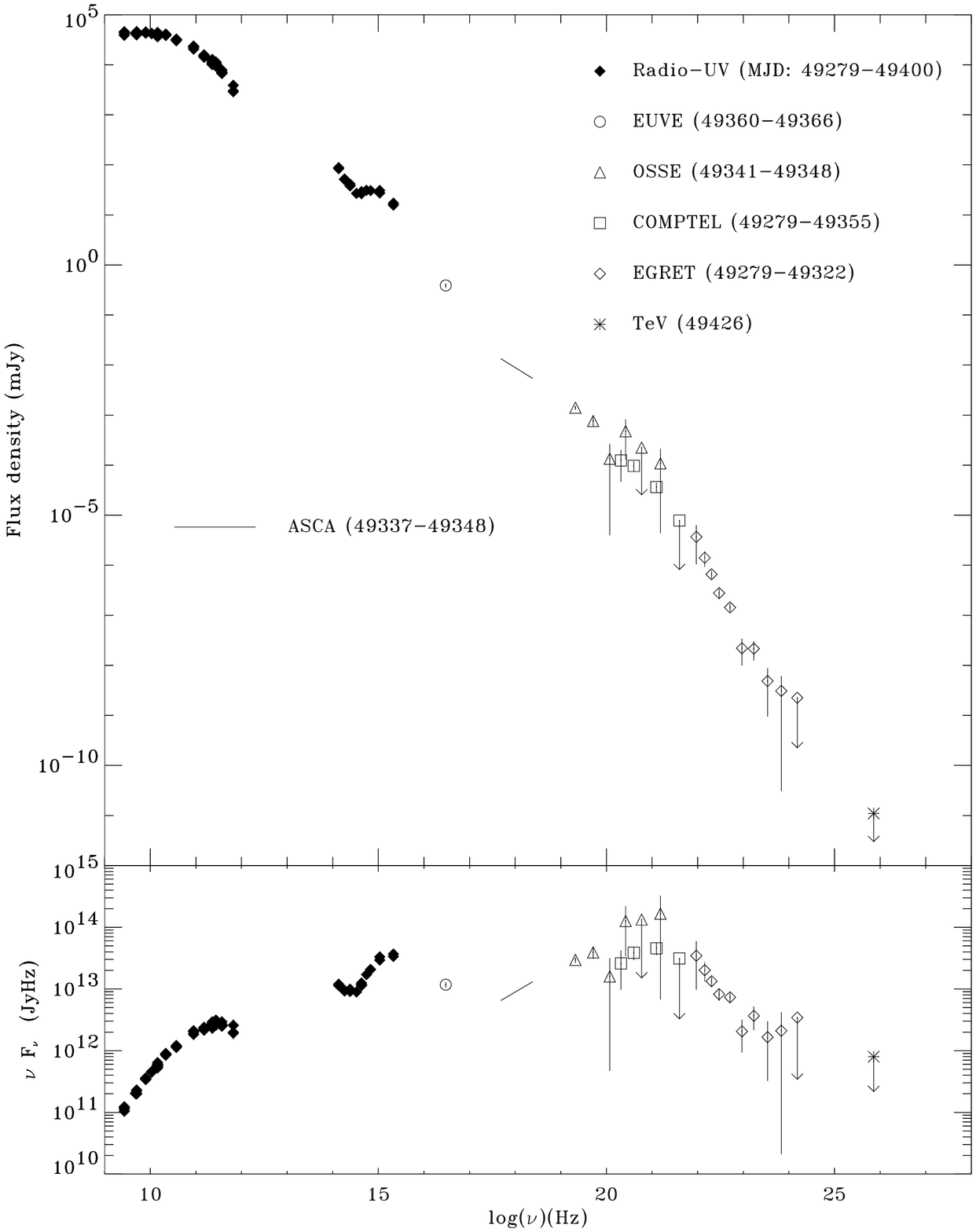] {Quasi-simultaneous
multiwavelength spectrum of 3C~273. The data in the radio through UV
range have not been averaged. \label{fig:mw_spectrum}}

\figcaption[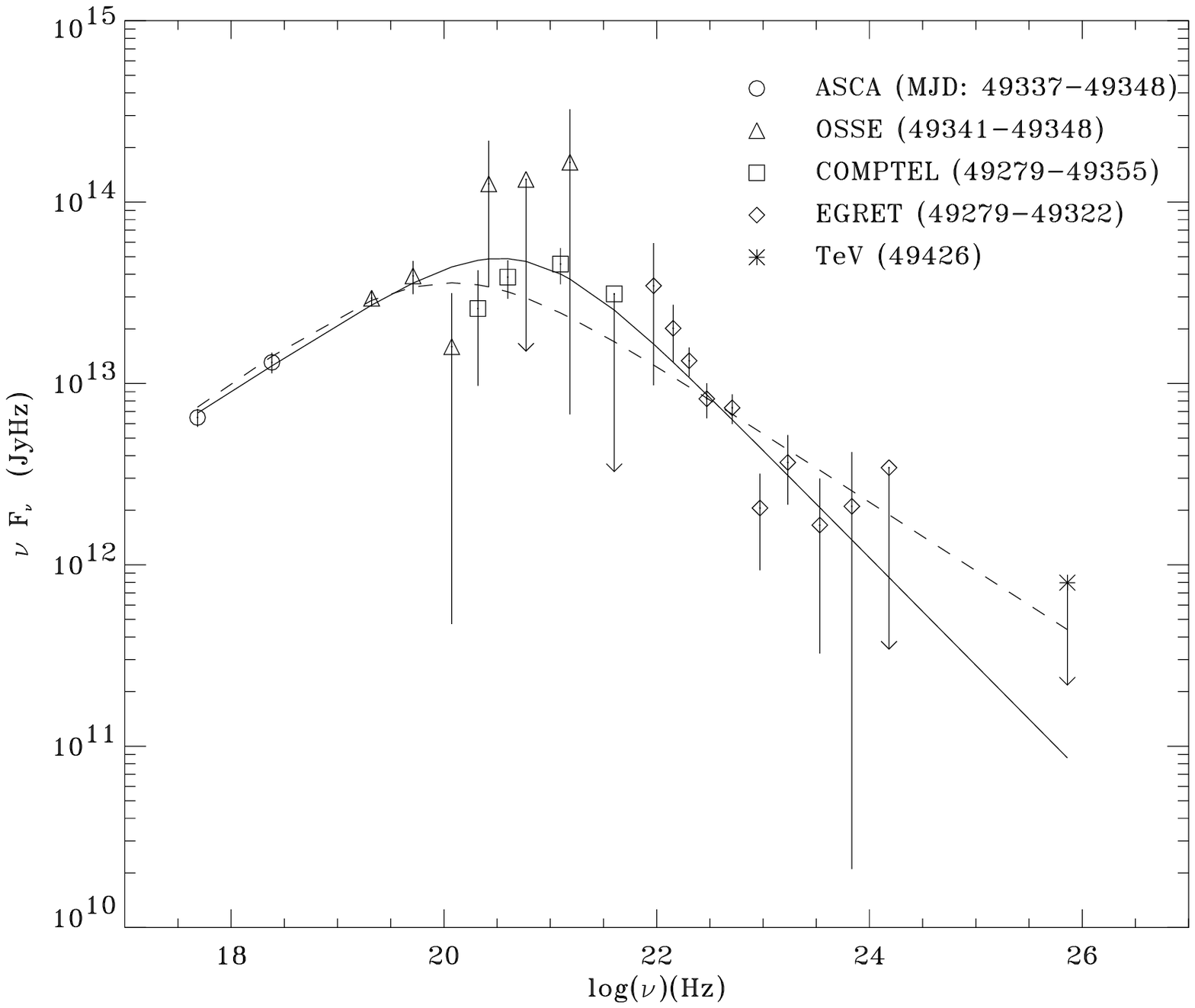] {High energy spectrum of
\3c. The solid line is the result from the fit in this paper and the
dashed line is the fit from Lichti \etal (1995). \label{fig:break}} 

\figcaption[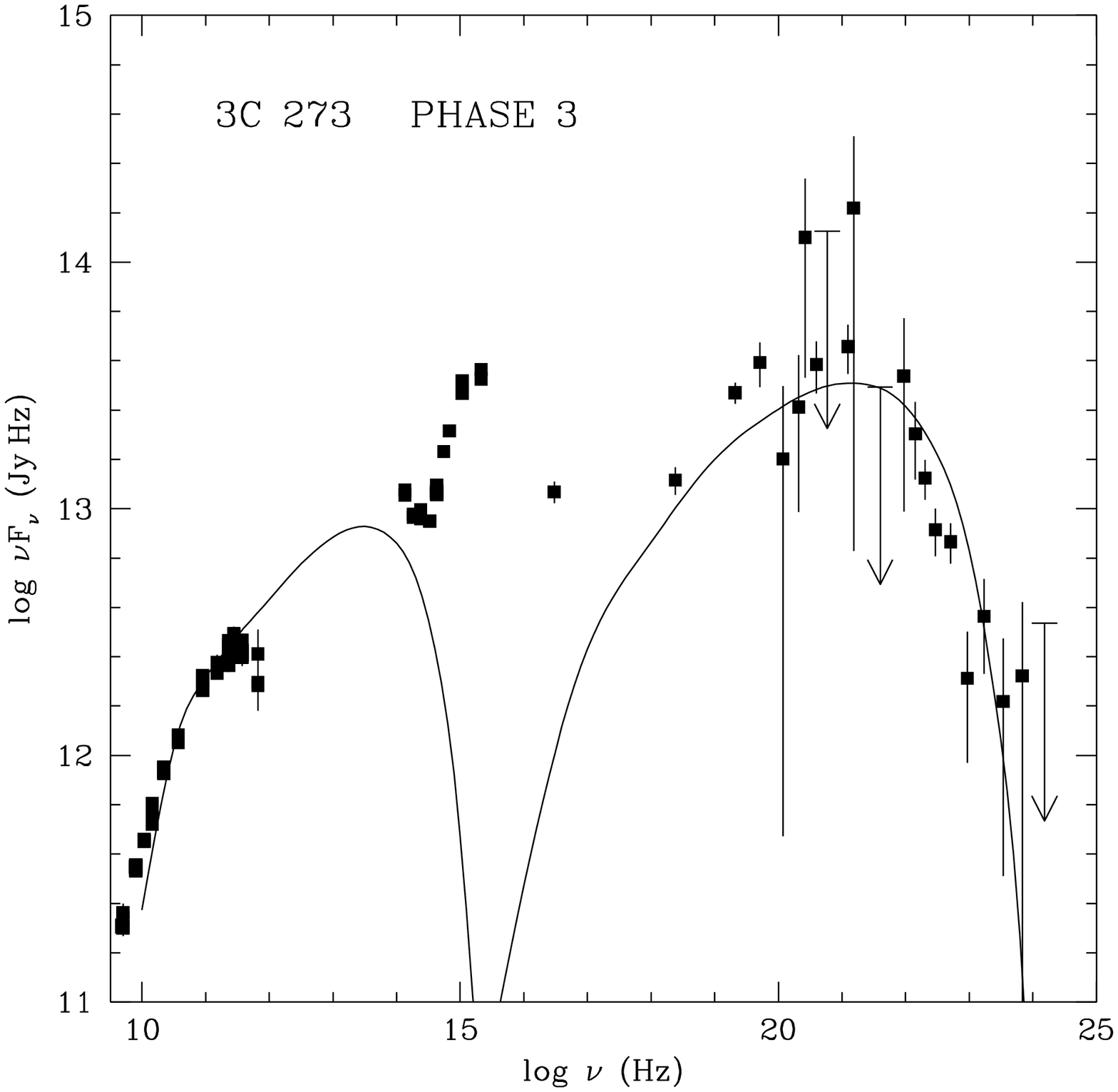] {Fit of the SSC-model to the
observed spectrum of \3c. Parameters of the model are: cross-sectional
radius r = 0.055 pc, opening half-angle of the jet: \deg{0.5}, angle
between the jet axis and the line of sight: \deg{6}, bulk Lorentz
factor: 9.3, minimum injected electron energy
\g$_{min}$ = 100, maximum injected electron energy  \g$_{max}$ = 2.5\c10{4},
injected power-law of electron energy distribution: 2.3, density of
relativistic electrons N$_{\rm e}$ = 180 cm$^{-3}$, magnetic field B =
0.023 G, ratio of randomly oriented to axial component of magnetic field:
1.5. \label{fig:SSC_fit}}

\figcaption[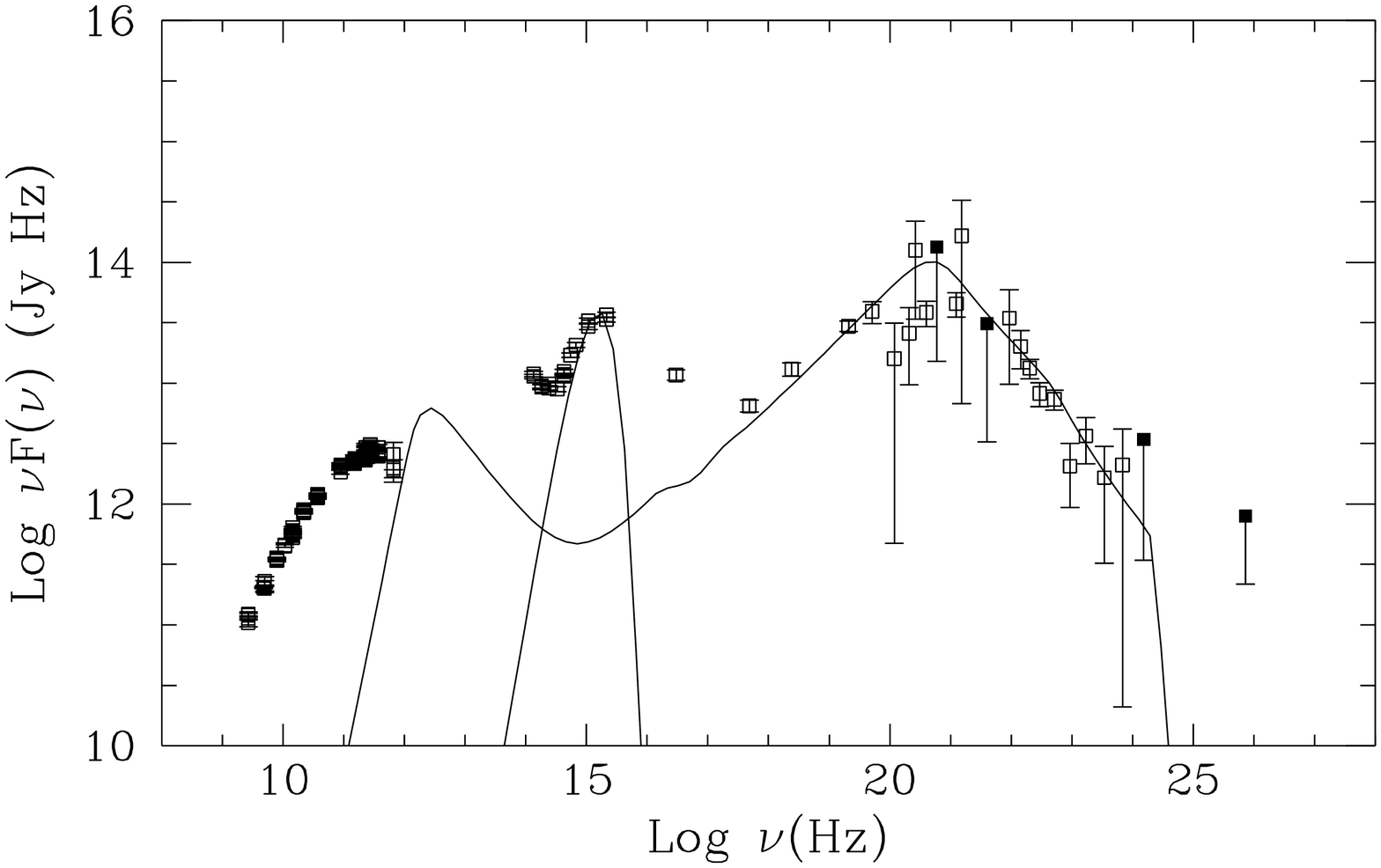] {Fit of the EC-model to
the observed spectrum of \3c. Parameters of the model are: Dimension
of the source R = 2\c10{16} cm, magnetic field B = 5.9 G, injected
luminosity L$_i$ = 3.7\c10{44} erg/s, beaming factor D = 6.5, U$_{ext}$'/U$_B$
= 2. \label{fig:EC_fit}}

\figcaption[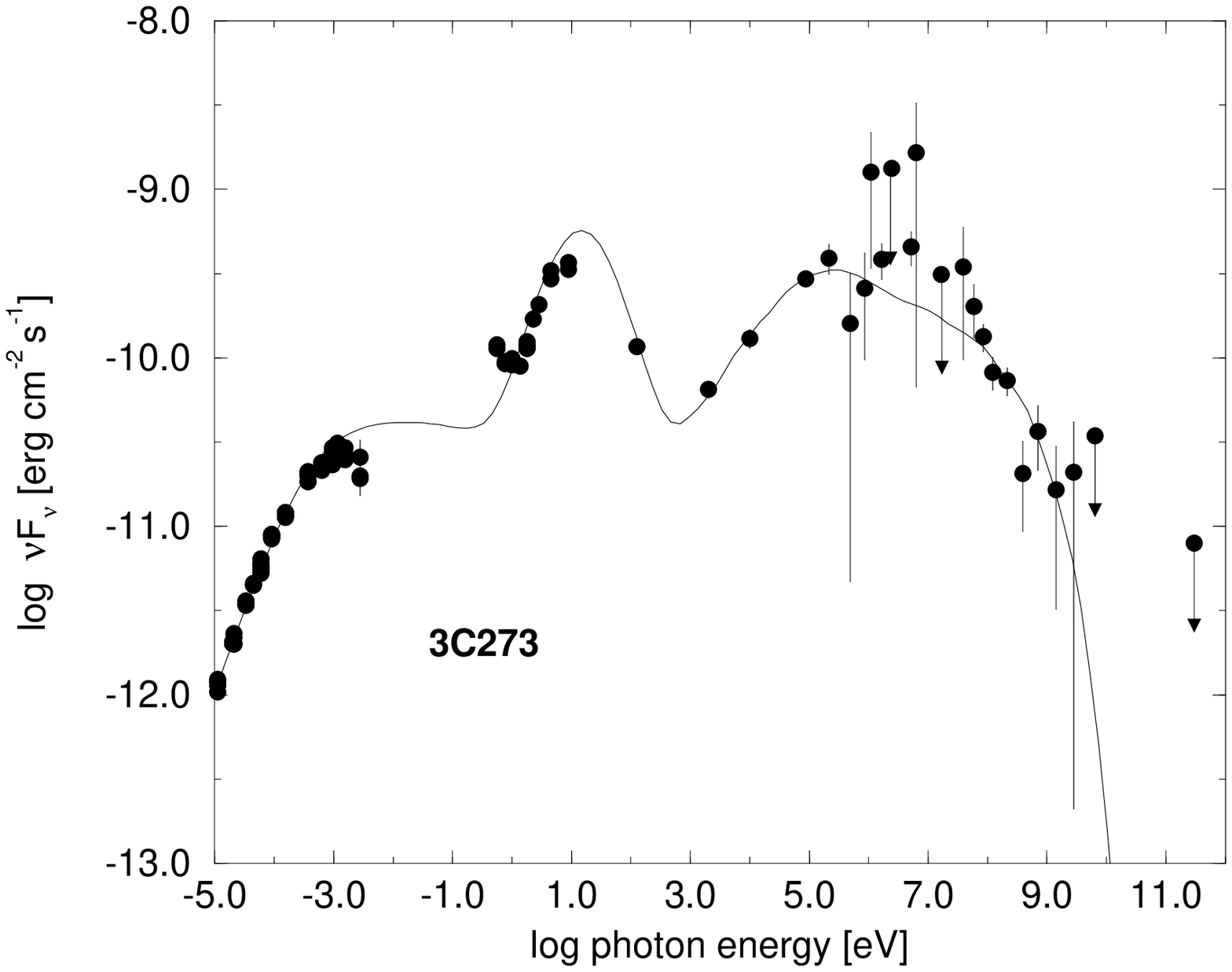] {Fit of the PIC-model to the
observed spectrum of \3c. Parameters of the model are: Jet radius
$r_\perp =$ \10{17}cm, jet Lorentz factor \g$_j =$ 8,
angle to the line of sight $\theta  =$ \deg{7}, relativistic
particle luminosity L$_j =$ 1.4\c10{46}erg/s, proton to electron
ratio $\eta =$ 100, equipartition magnetic field strength
B'$_{eq} =$ 0.8 G, ratio of proton and electron cooling rates 
$\xi =$ 0.15, maximum proton Lorentz factor $\gamma_{\rm p,max} = 
3\times 10^{10}$. \label{fig:PIC_fit}}

\clearpage
\plotone{f1.eps} 

\clearpage
\plotone{f2.eps} 

\clearpage
\plotone{f3.eps}

\clearpage
\plotone{f4.eps} 

\clearpage
\plotone{f5.eps} 

\clearpage
\plotone{f6.eps} 

\clearpage
\plotone{f7.eps} 

\clearpage
\plotone{f8.eps} 

\clearpage
\plotone{f9.eps} 

\begin{planotable}{lccccc}
\tablewidth{0pt}
\tablecaption{EGRET fluxes of \3c from phases 1, 2, 3, and cycle 4.}
\tablehead{ 
\colhead{Viewing}&
\colhead{Time of Observation}&
\colhead{MJD}&
\colhead{Offset}&
\colhead{Signif. }&
\colhead{Flux(E$>$100MeV)}\\
\colhead{Period}&
\colhead{}&
\colhead{}&
\colhead{(deg)}&
\colhead{$\sqrt{TS}$}&
\colhead{(10$^{-8}$\pcm2sec)}
}
\startdata
  3.0   & 06/15/91  -  06/28/91  &48422-48435&  4.29& 6.7  & 23.8\+-4.5\nl
 11.0   & 10/03/91  -  10/17/91  &48532-48546&  2.00& 2.5  & $<$ 17.7\nl
\hline
204.0   & 12/22/92  -  12/29/92  &48978-48685&  3.28& 0.5  & $<$ 18.2\nl
205.0   & 12/29/92  -  01/05/93  &48685-48992&  3.46& 2.3  & $<$ 33.2\nl
206.0   & 01/05/93  -  01/12/93  &48992-48999&  3.28& 2.5  & $<$ 46.5\nl
\hline
204.0 ... 206.0  & 12/22/92  -  01/12/93  &48978-48999& \nodata & 0.8  & $<$ 12.5\nl
\hline
304.0   & 10/19/93  -  10/25/93  &49279-49285&  5.38& 3.0  & $<$ 44.0\nl
305.0   & 10/25/93  -  11/02/93  &49285-49293&  5.69& 3.4  & 20.7\+- 8.1\nl
306.0   & 11/02/93  -  11/09/93  &49293-49300&  8.14& 2.8  & $<$ 42.0\nl
\hline
304.0 ... 306.0  & 10/19/93  -  11/09/93  &49279-49300& \nodata & 5.4 & 22.4\+- 5.3\nl
\hline
307.0   & 11/09/93  -  11/16/93  &49300-49307&  9.62& 4.1  & 35.1\+- 12.0\nl
308.0   & 11/16/93  -  11/19/93  &49307-49310& 10.54& 2.3  & $<$ 68.0\nl
\hline
307.0 ... 308.0  & 11/09/93  -  11/19/93  &49300-49310& \nodata & 4.7 & 34.0\+- 9.9\nl
\hline
308.6   & 11/23/93  -  12/01/93  &49314-49322& 10.54& 6.7  & 55.7\+- 11.9\nl
311.0   & 12/13/93  -  12/15/93  &49334-49336& 10.37& 0.0  & $<$ 34.0\nl
311.6   & 12/17/93  -  12/20/93  &49338-49341& 10.37& 1.7  & $<$ 61.0\nl
\hline
311.0 ... 311.6  & 12/13/93  -  12/20/93  &49334-49341& \nodata & 1.2  & $<$ 36.0\nl
\hline
312.0   & 12/20/93  -  12/27/93  &49341-49348&  7.28& 0.4  & $<$ 21.6\nl
313.0   & 12/27/93  -  01/03/94  &49348-49355& 14.34& 0.2  & $<$ 23.7\nl
\hline
311.6 ... 312.0  & 12/17/93  -  12/27/93  &49338-49348& \nodata & 0.1  & $<$ 18.9\nl
311.0 ... 313.0  & 12/13/93  -  01/03/94  &49334-49355& \nodata & 0.4  & $<$ 15.7\nl
\hline
405.0   & 11/29/94  -  12/07/94  &49685-49693& 11.30& 0.0  & $<$ 10.1\nl
406.0   & 12/13/94  -  12/20/94  &49699-49706& 18.77& 0.0  & $<$ 23.6\nl
407.0   & 12/20/94  -  01/03/95  &49706-49720& 19.33& 0.0  & $<$ 18.7\nl
408.0   & 01/03/95  -  01/10/95  &49720-49727& 10.36& 1.6  & $<$ 26.9\nl 
\enddata
\end{planotable}
 
\begin{planotable}{lcccccc}
\tablewidth{0pt}
\tablecaption{Results of spectral analysis.}
\tablehead{ 
\colhead{Viewing }&
\colhead{ Spectral     }&
\colhead{       N$_o$                }&
\colhead{ E$_o$ }&
\colhead{ $\chi^2/n_f$} \\
\colhead{Period  }&
\colhead{ Index (\Gg ) }&
\colhead{ \10{-9}\pcm2sec MeV$^{-1}$ }&
\colhead{  MeV  }&
\colhead{ } 
}
\startdata
305.0           & 3.20\+- 0.54 & 4.62\+- 1.23 & 102.7 & 0.71 \nl
304.0 ... 306.0 & 2.93\+- 0.33 & 1.75\+- 0.36 & 123.0 & 0.46 \nl
307.0           & 2.83\+- 0.35 & 6.83\+- 1.41 & 110.1 & 0.35 \nl
307.0 ... 308.0 & 2.69\+- 0.32 & 6.84\+- 1.33 & 118.4 & 0.53 \nl
308.6           & 2.20\+- 0.22 & 0.96\+- 0.19 & 225.7 & 0.63 \nl
304.0 ... 308.6 & 2.59\+- 0.15 & 1.58\+- 0.18 & 155.3 & 0.75 \nl
\enddata
\end{planotable}

\begin{planotable}{lcccccc}
\tablenum{3a}
\tablecolumns{7}
\tablewidth{0pt}
\tablecaption{2.7, 4.75 and 10.55 GHz radio data for \3c (Pohl, Reich).}
\tablehead{ 
\colhead{MJD}&
\multicolumn{2}{c}{  2.7~GHz }&
\multicolumn{2}{c}{  4.75~GHz}&
\multicolumn{2}{c}{  10.55~GHz} \\
\colhead{}&
\colhead{Flux(Jy)}&
\colhead{$\sigma$(Jy)}&
\colhead{Flux(Jy)}&
\colhead{$\sigma$(Jy)}&
\colhead{Flux(Jy)}&
\colhead{$\sigma$(Jy)}
}
\startdata
48335 & \nodata   &  \nodata  & 33.40 &  0.50 & \nodata   &  \nodata  \nl
48536 & 42.10 &  3.00 & 33.90 &  0.70 & 44.00 &  2.50 \nl
48567 & 37.10 &  0.60 & 34.00 &  0.70 & \nodata   &  \nodata  \nl
48588 & 39.40 &  1.70 & 35.20 &  1.20 & 45.70 &  1.20 \nl
48602 & 41.50 &  1.00 & 35.60 &  0.40 & 43.50 &  2.20 \nl
48673 & 40.00 &  2.70 & 36.50 &  1.50 & 44.90 &  3.00 \nl
48685 & 38.70 &  2.90 & \nodata   &  \nodata  & 45.70 &  4.80 \nl
48726 & 39.30 &  1.00 & \nodata   &  \nodata  & 47.10 &  2.50 \nl
48741 & 38.90 &  1.10 & 36.70 &  1.30 & 46.90 &  2.90 \nl
48775 & 41.70 &  1.60 & 39.80 &  1.80 & 46.50 &  2.30 \nl
48799 & 40.40 &  2.20 & \nodata   &  \nodata  & 43.30 &  4.00 \nl
48819 & \nodata   &  \nodata  & \nodata   &  \nodata  & 47.00 &  1.00 \nl
48883 & \nodata   &  \nodata  & \nodata   &  \nodata  & 45.00 &  1.00 \nl
48904 & 40.80 &  1.50 & \nodata   &  \nodata  & 44.80 &  1.00 \nl
48966 & 40.00 &  1.30 & 40.60 &  1.50 & 43.00 &  2.00 \nl
48967 & \nodata   &  \nodata  & \nodata   &  \nodata  & 44.80 &  2.00 \nl
48996 & 41.70 &  1.80 & 42.00 &  1.40 & 43.90 &  1.00 \nl
48997 & 42.20 &  1.60 & 41.60 &  1.40 & 45.60 &  2.20 \nl
49027 & 41.90 &  1.60 & 41.90 &  2.20 & 44.80 &  2.70 \nl
49056 & \nodata   &  \nodata  & \nodata   &  \nodata  & 43.40 &  2.20 \nl
49071 & 43.20 &  1.70 & 42.40 &  1.90 & 43.00 &  2.40 \nl
49108 & 42.30 &  1.70 & 43.00 &  2.30 & 41.10 &  1.50 \nl
49158 & \nodata   &  \nodata  & 43.20 &  1.30 & 41.70 &  1.50 \nl
49184 & 42.80 &  1.10 & \nodata   &  \nodata  & \nodata   &  \nodata  \nl
49185 & 42.80 &  1.20 & 43.20 &  1.80 & 41.40 &  1.50 \nl
49217 & 43.60 &  1.50 & 43.80 &  1.80 & \nodata   &  \nodata  \nl
49218 & 43.00 &  1.50 & \nodata   &  \nodata  & \nodata   &  \nodata  \nl
49219 & 45.90 &  1.70 & \nodata   &  \nodata  & 40.50 &  2.00 \nl
49297 & 45.70 &  2.10 & 41.60 &  1.70 & 40.90 &  2.70 \nl
49299 & \nodata   &  \nodata  & 43.10 &  1.70 & 39.30 &  2.10 \nl
49332 & 44.80 &  1.40 & \nodata   &  \nodata  & \nodata   &  \nodata  \nl
49349 & 44.80 &  1.70 & 42.70 &  1.60 & 39.80 &  1.60 \nl
49381 & 45.00 &  1.40 & 42.30 &  1.20 & 40.30 &  1.90 \nl
49413 & 44.70 &  1.60 & 42.80 &  1.60 & 39.70 &  1.80 \nl
49430 & 44.30 &  1.60 & 43.40 &  1.60 & 41.50 &  2.00 \nl
49713 & 46.30 &  2.60 & 43.00 &  2.50 & \nodata   &  \nodata  \nl
49756 & 45.90 &  1.90 & 41.10 &  1.30 & \nodata   &  \nodata  \nl
\enddata
\end{planotable}
\begin{planotable}{lcc}
\tablewidth{0pt}
\tablewidth{0pt}
\tablenum{3b}
\tablecaption{4.8 GHz light curve for \3c (Aller \& Aller).}
\tablehead{ 
\colhead{MJD}&\colhead{Flux}&\colhead{$\sigma$}\\
\colhead{   }&\colhead{(Jy)}&\colhead{(Jy)}
}
\startdata
48929 &  41.20 &  0.34\nl
48938 &  41.49 &  0.49\nl
48950 &  41.58 &  0.50\nl
48952 &  41.65 &  0.25\nl
48968 &  41.54 &  0.34\nl
48990 &  41.51 &  0.44\nl
49010 &  42.23 &  0.30\nl
49052 &  42.43 &  0.36\nl
49053 &  42.04 &  0.43\nl
49066 &  41.93 &  0.80\nl
49078 &  42.50 &  0.40\nl
49083 &  43.06 &  0.41\nl
49127 &  43.57 &  0.28\nl
49128 &  43.17 &  0.28\nl
49132 &  43.13 &  0.40\nl
49139 &  43.73 &  0.22\nl
49148 &  43.77 &  0.41\nl
49156 &  44.16 &  0.41\nl
49159 &  43.78 &  0.42\nl
49168 &  43.80 &  0.52\nl
49176 &  43.13 &  0.38\nl
49177 &  44.67 &  0.48\nl
49192 &  42.91 &  0.64\nl
49198 &  43.92 &  0.34\nl
49205 &  44.32 &  0.53\nl
49206 &  43.54 &  0.31\nl
49215 &  43.65 &  0.38\nl
49222 &  42.94 &  0.48\nl
49295 &  42.73 &  0.23\nl
49318 &  42.69 &  0.25\nl
49326 &  42.81 &  0.32\nl
49336 &  42.12 &  0.26\nl
49352 &  42.04 &  0.32\nl
49354 &  42.53 &  0.23\nl
49365 &  42.62 &  0.31\nl
49377 &  42.15 &  0.24\nl
49400 &  42.86 &  0.32\nl
49416 &  42.59 &  0.30\nl
49423 &  42.85 &  0.24\nl
49440 &  42.22 &  0.28\nl
49448 &  42.43 &  0.40\nl
49454 &  42.32 &  0.21\nl
49467 &  42.09 &  0.19\nl
49471 &  43.01 &  0.25\nl
49482 &  42.99 &  0.25\nl
49498 &  42.86 &  0.33\nl
49516 &  43.42 &  0.29\nl
49519 &  43.50 &  0.19\nl
49520 &  43.02 &  0.25\nl
49531 &  43.41 &  0.25\nl
49569 &  43.25 &  0.27\nl
49575 &  42.81 &  0.20\nl
49583 &  42.67 &  0.37\nl
49737 &  41.18 &  0.69\nl
49769 &  41.80 &  0.82\nl
49783 &  43.28 &  1.06\nl
\enddata
\end{planotable}
\begin{planotable}{lcc}
\tablewidth{0pt}
\tablenum{3c}
\tablecaption{8 GHz light curve for \3c (Aller \& Aller).}
\tablehead{ 
\colhead{MJD}&\colhead{Flux}&\colhead{$\sigma$}\\
\colhead{   }&\colhead{(Jy)}&\colhead{(Jy)}
}
\startdata
 48927 &  49.59 &  0.58\nl
 48935 &  46.48 &  0.44\nl
 48951 &  47.25 &  0.47\nl
 48955 &  46.98 &  0.31\nl
 48969 &  44.50 &  0.44\nl
 48981 &  47.60 &  0.48\nl
 48982 &  47.32 &  0.42\nl
 48988 &  47.47 &  0.35\nl
 48994 &  47.15 &  0.40\nl
 49001 &  43.51 &  0.49\nl
 49007 &  48.31 &  0.45\nl
 49019 &  50.39 &  0.80\nl
 49020 &  47.38 &  0.65\nl
 49037 &  47.32 &  0.34\nl
 49058 &  46.38 &  0.43\nl
 49071 &  46.59 &  0.48\nl
 49086 &  48.22 &  0.54\nl
 49114 &  47.03 &  0.51\nl
 49122 &  47.06 &  0.56\nl
 49150 &  46.01 &  0.60\nl
 49152 &  46.69 &  0.46\nl
 49161 &  47.14 &  0.57\nl
 49168 &  48.38 &  0.55\nl
 49178 &  47.47 &  0.67\nl
 49190 &  45.92 &  0.44\nl
 49200 &  46.51 &  0.39\nl
 49207 &  46.60 &  0.48\nl
 49209 &  46.90 &  0.45\nl
 49217 &  45.84 &  0.54\nl
 49226 &  47.12 &  0.43\nl
 49237 &  46.55 &  0.65\nl
 49243 &  45.03 &  0.47\nl
 49294 &  44.28 &  0.51\nl
 49301 &  45.10 &  0.33\nl
 49305 &  44.14 &  0.44\nl
 49322 &  42.85 &  0.78\nl
 49334 &  44.12 &  0.72\nl
 49346 &  44.53 &  0.50\nl
 49347 &  44.45 &  0.47\nl
 49360 &  43.20 &  0.46\nl
 49368 &  45.24 &  0.37\nl
 49385 &  44.86 &  0.66\nl
 49414 &  44.15 &  0.79\nl
 49415 &  45.10 &  0.81\nl
 49421 &  45.79 &  0.63\nl
 49430 &  43.47 &  0.51\nl
 49431 &  44.00 &  0.62\nl
 49434 &  45.33 &  0.41\nl
 49447 &  45.46 &  0.33\nl
 49453 &  44.85 &  0.49\nl
 49461 &  44.58 &  0.53\nl
 49462 &  45.15 &  0.53\nl
 49476 &  44.17 &  0.44\nl
 49491 &  44.40 &  0.43\nl
 49506 &  43.84 &  0.51\nl
 49514 &  42.59 &  0.82\nl
 49518 &  42.49 &  0.55\nl
 49539 &  43.32 &  0.65\nl
 49561 &  43.55 &  0.36\nl
 49572 &  42.39 &  0.34\nl
 49573 &  43.29 &  0.48\nl
 49586 &  42.27 &  0.39\nl
 49588 &  41.93 &  0.85\nl
 49589 &  45.45 &  0.87\nl
 49593 &  44.02 &  0.52\nl
 49594 &  42.06 &  0.50\nl
 49595 &  41.56 &  0.51\nl
 49650 &  42.43 &  0.39\nl
 49651 &  42.16 &  0.34\nl
 49670 &  42.39 &  0.48\nl
 49672 &  42.36 &  0.37\nl
 49689 &  41.61 &  0.73\nl
 49705 &  40.09 &  0.86\nl
 49708 &  40.70 &  0.57\nl
 49709 &  40.08 &  0.68\nl
 49710 &  41.36 &  0.35\nl
 49711 &  42.38 &  0.48\nl
 49712 &  40.46 &  0.43\nl
 49736 &  39.38 &  0.41\nl
 49753 &  41.11 &  0.34\nl
 49756 &  39.92 &  0.31\nl
 49762 &  40.57 &  0.45\nl
 49765 &  40.66 &  0.33\nl
 49766 &  39.99 &  0.40\nl
 49780 &  40.70 &  0.29\nl
 49781 &  40.00 &  0.27\nl
\enddata
\end{planotable}

\begin{planotable}{lcc}
\tablewidth{0pt}
\tablenum{3d}
\tablecaption{14.5 GHz light curve for \3c (Aller \& Aller).}
\tablehead{ 
\colhead{MJD}&\colhead{Flux}&\colhead{$\sigma$}\\
\colhead{   }&\colhead{(Jy)}&\colhead{(Jy)}
}
\startdata
 48930 &  45.36 &  0.30\nl
 48937 &  45.52 &  0.32\nl
 48946 &  46.28 &  0.57\nl
 48963 &  45.54 &  0.40\nl
 48975 &  44.95 &  0.61\nl
 48979 &  45.72 &  0.35\nl
 48986 &  45.71 &  0.44\nl
 48988 &  46.83 &  0.67\nl
 49048 &  43.00 &  0.35\nl
 49050 &  43.40 &  0.53\nl
 49059 &  42.78 &  0.49\nl
 49068 &  40.86 &  0.67\nl
 49090 &  40.32 &  0.38\nl
 49106 &  39.35 &  0.46\nl
 49109 &  36.65 &  0.52\nl
 49124 &  43.54 &  0.50\nl
 49130 &  37.66 &  0.48\nl
 49142 &  41.87 &  1.02\nl
 49152 &  38.34 &  0.39\nl
 49165 &  39.49 &  0.53\nl
 49175 &  38.45 &  0.81\nl
 49185 &  40.42 &  0.81\nl
 49194 &  39.31 &  0.79\nl
 49202 &  40.91 &  1.06\nl
 49213 &  39.64 &  0.93\nl
 49220 &  37.27 &  0.47\nl
 49231 &  35.05 &  0.70\nl
 49275 &  36.75 &  1.04\nl
 49292 &  37.84 &  1.10\nl
 49300 &  39.41 &  0.49\nl
 49310 &  38.69 &  0.69\nl
 49311 &  39.47 &  0.34\nl
 49312 &  36.38 &  0.57\nl
 49329 &  40.19 &  0.48\nl
 49343 &  41.14 &  0.36\nl
 49350 &  40.43 &  0.69\nl
 49363 &  44.25 &  0.75\nl
 49370 &  41.77 &  0.71\nl
 49393 &  41.07 &  0.34\nl
 49412 &  41.67 &  0.39\nl
 49417 &  41.49 &  0.67\nl
 49419 &  41.84 &  0.38\nl
 49441 &  39.09 &  0.29\nl
 49450 &  40.22 &  0.41\nl
 49458 &  39.94 &  1.13\nl
 49475 &  38.38 &  0.44\nl
 49478 &  40.27 &  0.79\nl
 49486 &  38.25 &  0.78\nl
 49503 &  37.83 &  0.42\nl
 49505 &  36.85 &  0.28\nl
 49508 &  37.11 &  0.53\nl
 49512 &  37.22 &  0.63\nl
 49513 &  36.88 &  0.97\nl
 49535 &  34.40 &  0.73\nl
 49547 &  35.63 &  0.28\nl
 49551 &  35.40 &  0.48\nl
 49564 &  36.80 &  0.43\nl
 49604 &  32.65 &  0.74\nl
 49654 &  33.34 &  0.77\nl
 49664 &  33.63 &  0.34\nl
 49714 &  34.59 &  0.47\nl
 49743 &  33.00 &  0.38\nl
 49768 &  32.10 &  0.35\nl
 49777 &  32.32 &  0.32\nl
 49779 &  32.32 &  0.40\nl
\enddata
\end{planotable}

\begin{planotable}{lcc}
\tablewidth{0pt}
\tablenum{3e}
\tablecaption{22 GHz light curve for \3c (Ter\"asranta, Tornikoski, Valtaoja).}
\tablehead{ 
\colhead{MJD}&\colhead{Flux}&\colhead{$\sigma$}\\
\colhead{   }&\colhead{(Jy)}&\colhead{(Jy)}
}
\startdata
 48305 &  38.62 &  0.79\nl
 48315 &  38.16 &  0.65\nl
 48324 &  40.11 &  0.63\nl
 48357 &  41.84 &  0.64\nl
 48371 &  42.45 &  0.59\nl
 48391 &  45.34 &  0.68\nl
 48410 &  47.34 &  1.48\nl
 48418 &  51.47 &  1.33\nl
 48438 &  49.58 &  1.49\nl
 48449 &  48.95 &  1.56\nl
 48459 &  52.22 &  1.54\nl
 48466 &  53.07 &  1.36\nl
 48473 &  55.58 &  1.66\nl
 48483 &  56.09 &  2.37\nl
 48494 &  55.37 &  1.68\nl
 48501 &  55.98 &  1.33\nl
 48555 &  53.36 &  0.98\nl
 48583 &  50.37 &  1.09\nl
 48593 &  51.58 &  1.37\nl
 48602 &  50.97 &  0.83\nl
 48624 &  50.60 &  0.97\nl
 48631 &  49.55 &  0.96\nl
 48642 &  48.46 &  0.62\nl
 48648 &  47.83 &  1.02\nl
 48661 &  49.24 &  1.11\nl
 48681 &  46.72 &  0.80\nl
 48721 &  44.19 &  1.30\nl
 48728 &  45.83 &  0.76\nl
 48737 &  43.00 &  0.78\nl
 48745 &  43.46 &  1.13\nl
 48780 &  42.67 &  1.76\nl
 48800 &  43.00 &  1.52\nl
 48805 &  42.99 &  1.87\nl
 48812 &  45.29 &  1.93\nl
 48825 &  42.15 &  0.88\nl
 48830 &  41.99 &  1.99\nl
 48863 &  42.69 &  1.02\nl
 48917 &  40.66 &  0.67\nl
 48927 &  41.32 &  1.12\nl
 48967 &  42.06 &  0.70\nl
 48977 &  40.46 &  0.68\nl
 48985 &  40.33 &  0.83\nl
 48993 &  40.69 &  0.78\nl
 49026 &  37.53 &  0.98\nl
 49041 &  36.80 &  0.60\nl
 49051 &  36.25 &  0.66\nl
 49058 &  35.06 &  0.82\nl
 49098 &  33.68 &  0.64\nl
 49107 &  35.35 &  1.09\nl
 49118 &  36.38 &  1.08\nl
 49125 &  34.86 &  1.12\nl
 49134 &  34.74 &  0.61\nl
 49172 &  37.17 &  1.08\nl
 49178 &  37.80 &  1.61\nl
 49199 &  37.62 &  1.09\nl
 49211 &  34.56 &  1.09\nl
 49224 &  35.40 &  1.57\nl
 49284 &  39.18 &  1.46\nl
 49293 &  38.50 &  0.50\nl
 49301 &  40.94 &  0.86\nl
 49309 &  40.34 &  0.48\nl
 49322 &  39.36 &  0.49\nl
 49335 &  41.07 &  0.63\nl
 49345 &  40.53 &  0.63\nl
 49423 &  36.84 &  0.49\nl
 49429 &  36.84 &  0.59\nl
 49446 &  33.51 &  0.74\nl
 49643 &  29.56 &  0.65\nl
 49661 &  29.69 &  0.84\nl
 49671 &  30.49 &  0.50\nl
 49678 &  29.96 &  0.65\nl
 49703 &  32.21 &  0.83\nl
 49713 &  30.44 &  0.38\nl
 49722 &  31.31 &  0.40\nl
 49736 &  31.17 &  0.55\nl
 49745 &  30.20 &  0.36\nl
 49753 &  29.58 &  0.63\nl
 49759 &  29.00 &  0.60\nl
 49814 &  27.42 &  0.42\nl
 49825 &  27.14 &  0.58\nl
 49831 &  27.30 &  0.66\nl
 49840 &  27.39 &  0.89\nl
\enddata
\end{planotable}
\begin{planotable}{lcc}
\tablewidth{0pt}
\tablenum{3f}
\tablecaption{37 GHz light curve for \3c (Ter\"asranta, Tornikoski, Valtaoja).}
\tablehead{ 
\colhead{MJD}&\colhead{Flux}&\colhead{$\sigma$}\\
\colhead{   }&\colhead{(Jy)}&\colhead{(Jy)}
}
\startdata
 48308 &  36.88 &  0.58\nl
 48319 &  37.54 &  0.91\nl
 48343 &  39.71 &  0.41\nl
 48350 &  39.96 &  0.66\nl
 48367 &  41.10 &  0.39\nl
 48373 &  42.48 &  0.65\nl
 48389 &  44.07 &  0.67\nl
 48402 &  46.53 &  0.95\nl
 48413 &  47.27 &  0.60\nl
 48420 &  47.03 &  0.94\nl
 48437 &  46.23 &  0.77\nl
 48450 &  47.77 &  0.57\nl
 48457 &  49.33 &  0.89\nl
 48468 &  52.19 &  0.64\nl
 48473 &  51.80 &  1.08\nl
 48482 &  55.92 &  1.22\nl
 48493 &  55.70 &  0.62\nl
 48500 &  54.27 &  1.34\nl
 48555 &  49.64 &  0.77\nl
 48573 &  45.40 &  0.99\nl
 48581 &  44.83 &  1.03\nl
 48590 &  49.58 &  1.03\nl
 48607 &  45.78 &  0.94\nl
 48627 &  42.59 &  0.91\nl
 48642 &  42.05 &  0.51\nl
 48652 &  41.48 &  0.50\nl
 48665 &  41.00 &  0.86\nl
 48673 &  40.21 &  0.83\nl
 48722 &  37.57 &  0.53\nl
 48733 &  37.24 &  0.48\nl
 48741 &  37.44 &  0.65\nl
 48745 &  38.21 &  0.85\nl
 48777 &  36.42 &  0.38\nl
 48796 &  37.14 &  0.60\nl
 48801 &  36.87 &  0.80\nl
 48817 &  36.60 &  0.49\nl
 48826 &  36.06 &  0.61\nl
 48868 &  35.56 &  0.52\nl
 48914 &  34.56 &  0.78\nl
 48924 &  33.45 &  0.44\nl
 48977 &  32.03 &  0.78\nl
 48985 &  31.76 &  0.38\nl
 48993 &  32.01 &  0.40\nl
 49026 &  28.95 &  0.40\nl
 49031 &  29.45 &  0.72\nl
 49043 &  27.23 &  0.52\nl
 49047 &  28.32 &  0.61\nl
 49057 &  27.73 &  0.43\nl
 49064 &  27.51 &  0.41\nl
 49096 &  26.64 &  0.46\nl
 49104 &  26.96 &  0.37\nl
 49112 &  27.13 &  0.55\nl
 49119 &  27.25 &  0.44\nl
 49129 &  28.78 &  0.43\nl
 49134 &  29.37 &  0.63\nl
 49164 &  28.45 &  0.36\nl
 49172 &  27.69 &  0.66\nl
 49201 &  29.29 &  0.38\nl
 49211 &  30.92 &  0.51\nl
 49216 &  30.85 &  0.75\nl
 49225 &  29.56 &  0.75\nl
 49288 &  32.59 &  0.55\nl
 49295 &  31.82 &  0.89\nl
 49306 &  31.98 &  0.60\nl
 49314 &  30.91 &  0.78\nl
 49329 &  32.21 &  0.66\nl
 49346 &  31.41 &  0.52\nl
 49360 &  32.21 &  0.77\nl
 49385 &  30.29 &  0.80\nl
 49433 &  27.36 &  0.68\nl
 49445 &  26.16 &  0.42\nl
 49646 &  23.59 &  0.53\nl
 49664 &  25.22 &  1.17\nl
 49677 &  24.41 &  0.40\nl
 49712 &  24.83 &  0.51\nl
 49721 &  25.80 &  0.71\nl
 49734 &  24.92 &  0.55\nl
 49748 &  23.44 &  0.57\nl
 49758 &  23.30 &  0.30\nl
 49823 &  21.18 &  0.46\nl
 49832 &  20.47 &  0.55\nl
 49838 &  21.36 &  0.47\nl
\enddata
\end{planotable}

\begin{planotable}{lcc}
\tablewidth{0pt}
\tablenum{4a}
\tablecaption{0.45 mm fluxes of \3c (McHardy).}
\tablehead{ 
\colhead{MJD}&\colhead{Flux}&\colhead{$\sigma$}\\
\colhead{   }&\colhead{(Jy)}&\colhead{(Jy)}
}
\startdata
 49137 &   5.70 &  1.00\nl
 49139 &   5.20 &  1.00\nl
 49140 &   6.80 &  1.00\nl
 49166 &   5.80 &  1.20\nl
 49214 &   3.60 &  1.00\nl
 49342 &   3.00 &  0.50\nl
 49344 &   3.90 &  1.00\nl
 49347 &   2.90 &  0.60\nl
 49409 &   3.80 &  0.60\nl
 49413 &   4.00 &  0.80\nl
\enddata
\end{planotable}

\begin{planotable}{lcc}
\tablewidth{0pt}
\tablenum{4b}
\tablecaption{0.8 mm fluxes of \3c (McHardy).}
\tablehead{ 
\colhead{MJD}&\colhead{Flux}&\colhead{$\sigma$}\\
\colhead{   }&\colhead{(Jy)}&\colhead{(Jy)}
}
\startdata
 49118 &  12.20 &  1.50\nl
 49118 &  10.70 &  1.20\nl
 49119 &   8.80 &  1.70\nl
 49137 &  10.50 &  1.20\nl
 49139 &  11.70 &  1.20\nl
 49140 &  11.30 &  1.00\nl
 49161 &   8.20 &  0.60\nl
 49166 &  10.50 &  0.50\nl
 49214 &   9.70 &  0.70\nl
 49310 &   6.70 &  0.50\nl
 49311 &   7.00 &  0.50\nl
 49311 &   6.90 &  0.50\nl
 49312 &   6.90 &  0.50\nl
 49342 &   7.00 &  0.50\nl
 49343 &   7.10 &  0.50\nl
 49344 &   6.80 &  0.50\nl
 49347 &   6.90 &  0.40\nl
 49352 &   7.90 &  0.50\nl
 49353 &   7.00 &  0.50\nl
 49361 &   7.20 &  0.60\nl
 49409 &   7.70 &  0.50\nl
 49413 &   8.00 &  0.60\nl
 49422 &   9.00 &  0.50\nl
 49437 &   8.90 &  0.90\nl
 49450 &   8.30 &  0.60\nl
 49453 &  10.00 &  0.70\nl
 49465 &   7.80 &  0.50\nl
 49473 &   8.80 &  0.80\nl
\enddata
\end{planotable}

\begin{planotable}{lcc}
\tablewidth{0pt}
\tablenum{4c}
\tablecaption{ 1.1 mm fluxes of \3c (McHardy). }
\tablehead{ 
\colhead{MJD}&\colhead{Flux}&\colhead{$\sigma$}\\
\colhead{   }&\colhead{(Jy)}&\colhead{(Jy)}
}
\startdata
 49118 &  14.20 &  1.00\nl
 49118 &  14.10 &  1.00\nl
 49119 &  14.50 &  0.90\nl
 49123 &  14.70 &  1.00\nl
 49137 &  14.00 &  0.80\nl
 49139 &  14.80 &  0.80\nl
 49140 &  14.10 &  0.70\nl
 49161 &  11.80 &  1.00\nl
 49166 &  14.40 &  0.90\nl
 49183 &  16.74 &  1.50\nl
 49187 &  14.60 &  0.50\nl
 49214 &  12.60 &  0.80\nl
 49310 &   9.80 &  0.40\nl
 49311 &  10.00 &  0.60\nl
 49311 &   9.70 &  0.50\nl
 49312 &   9.40 &  0.50\nl
 49314 &  10.20 &  0.60\nl
 49315 &  10.10 &  0.50\nl
 49316 &  10.50 &  0.50\nl
 49318 &  10.20 &  0.40\nl
 49319 &  10.20 &  0.50\nl
 49320 &  10.10 &  0.90\nl
 49326 &   9.90 &  0.60\nl
 49329 &  10.20 &  0.70\nl
 49342 &   9.40 &  0.50\nl
 49343 &   9.80 &  0.50\nl
 49344 &   9.60 &  0.50\nl
 49347 &   9.90 &  0.50\nl
 49348 &  10.10 &  0.60\nl
 49349 &  10.20 &  0.70\nl
 49350 &  10.40 &  0.70\nl
 49351 &  10.80 &  0.60\nl
 49352 &  10.50 &  0.60\nl
 49353 &  10.30 &  0.50\nl
 49354 &  10.50 &  0.70\nl
 49355 &  10.50 &  0.50\nl
 49356 &  11.20 &  0.90\nl
 49357 &  11.30 &  0.80\nl
 49358 &  11.30 &  0.80\nl
 49359 &  11.30 &  0.70\nl
 49360 &  10.50 &  0.70\nl
 49361 &  10.30 &  0.60\nl
 49409 &  10.90 &  0.70\nl
 49413 &  10.50 &  0.70\nl
 49422 &  11.00 &  0.40\nl
 49437 &  10.80 &  0.80\nl
 49450 &  10.80 &  0.50\nl
 49453 &  12.10 &  0.70\nl
 49465 &  10.40 &  0.60\nl
 49472 &  10.10 &  0.60\nl
 49473 &  10.70 &  0.60\nl
 49474 &  10.70 &  0.60\nl
\enddata
\end{planotable}

\begin{planotable}{lcc}
\tablewidth{0pt}
\tablenum{4d}
\tablecaption{1.3 mm fluxes of \3c (Tornikoski, Ter\"asranta, Valtaoja, Marscher, McHardy, Robson).  }
\tablehead{ 
\colhead{MJD}&\colhead{Flux}&\colhead{$\sigma$}\\
\colhead{   }&\colhead{(Jy)}&\colhead{(Jy)}
}
\startdata
 48269 &  12.40 &  0.70\nl
 48300 &  17.60 &  1.10\nl
 48340 &  19.15 &  0.24\nl
 48341 &  18.13 &  0.10\nl
 48342 &  18.27 &  0.08\nl
 48351 &  22.26 &  1.60\nl
 48353 &  18.43 &  0.10\nl
 48353 &  18.47 &  0.09\nl
 48354 &  18.30 &  0.40\nl
 48354 &  18.51 &  0.06\nl
 48366 &  19.25 &  0.80\nl
 48393 &  17.96 &  0.90\nl
 48417 &  18.16 &  0.90\nl
 48441 &  13.38 &  1.03\nl
 48454 &  17.65 &  0.80\nl
 48480 &  20.60 &  1.20\nl
 48493 &  20.60 &  1.00\nl
 48514 &  19.60 &  1.00\nl
 48530 &  16.00 &  0.10\nl
 48533 &  15.10 &  0.10\nl
 48559 &  15.47 &  0.04\nl
 48571 &  14.53 &  0.19\nl
 48589 &  17.44 &  0.80\nl
 48607 &  16.90 &  0.07\nl
 48653 &  13.90 &  0.80\nl
 48654 &  12.60 &  1.00\nl
 48668 &  13.50 &  1.00\nl
 48697 &  14.60 &  0.90\nl
 48710 &  15.00 &  1.50\nl
 48714 &  13.60 &  1.20\nl
 48727 &  14.30 &  1.00\nl
 48749 &  13.50 &  0.40\nl
 48776 &  10.92 &  0.29\nl
 48777 &  11.49 &  0.48\nl
 48858 &  12.50 &  0.88\nl
 49012 &  10.10 &  0.50\nl
 49025 &  13.70 &  0.70\nl
 49030 &  12.90 &  1.00\nl
 49031 &  13.20 &  1.40\nl
 49032 &  13.00 &  1.00\nl
 49044 &  14.32 &  0.57\nl
 49045 &  15.50 &  1.10\nl
 49046 &  15.27 &  0.58\nl
 49057 &  13.80 &  1.00\nl
 49082 &  16.90 &  0.50\nl
 49093 &  16.90 &  0.90\nl
 49094 &  16.00 &  0.90\nl
 49118 &  15.20 &  1.00\nl
 49118 &  16.20 &  1.00\nl
 49119 &  13.00 &  1.50\nl
 49137 &  15.20 &  0.80\nl
 49139 &  15.90 &  0.80\nl
 49140 &  15.40 &  0.80\nl
 49157 &  14.55 &  0.82\nl
 49166 &  14.90 &  0.80\nl
 49214 &  13.30 &  0.70\nl
 49312 &  11.50 &  0.80\nl
 49315 &  10.90 &  0.60\nl
 49322 &  12.18 &  0.41\nl
 49342 &  10.10 &  0.60\nl
 49343 &  10.30 &  0.50\nl
 49344 &  10.20 &  0.50\nl
 49347 &  10.40 &  0.60\nl
 49352 &  10.60 &  0.60\nl
 49353 &  10.80 &  0.50\nl
 49354 &  11.30 &  0.80\nl
 49360 &  11.10 &  0.80\nl
 49361 &  10.60 &  0.60\nl
 49409 &  11.20 &  0.70\nl
 49413 &  12.30 &  0.90\nl
 49422 &  11.80 &  0.50\nl
 49453 &  13.10 &  0.80\nl
 49465 &  11.40 &  0.70\nl
 49472 &  10.30 &  0.60\nl
 49473 &  11.30 &  1.00\nl
 49474 &  11.00 &  0.90\nl
 49527 &  13.95 &  0.98\nl
 49527 &  13.99 &  0.98\nl
 49528 &  14.10 &  1.00\nl
 49528 &  14.97 &  1.05\nl
 49775 &   9.53 &  0.77\nl
 49775 &  10.37 &  0.83\nl
\enddata
\end{planotable}
\begin{planotable}{lcc}
\tablewidth{0pt}
\tablenum{4e}
\tablecaption{2.0 mm fluxes of \3c (McHardy).}
\tablehead{ 
\colhead{MJD}&\colhead{Flux}&\colhead{$\sigma$}\\
\colhead{   }&\colhead{(Jy)}&\colhead{(Jy)}
}
\startdata
 49118 &  22.60 &  1.70 \nl
 49118 &  20.90 &  1.50 \nl
 49119 &  20.70 &  1.20 \nl
 49137 &  19.60 &  1.00 \nl
 49139 &  19.70 &  1.00 \nl
 49140 &  20.20 &  1.00 \nl
 49166 &  18.90 &  1.00 \nl
 49214 &  18.20 &  0.50 \nl
 49310 &  15.30 &  1.10 \nl
 49311 &  15.70 &  1.10 \nl
 49312 &  15.10 &  0.70 \nl
 49314 &  15.70 &  1.00 \nl
 49315 &  15.20 &  0.90 \nl
 49316 &  15.00 &  0.80 \nl
 49318 &  15.20 &  0.90 \nl
 49319 &  15.20 &  0.90 \nl
 49326 &  14.60 &  0.80 \nl
 49329 &  15.70 &  1.20 \nl
 49342 &  14.40 &  0.80 \nl
 49343 &  14.30 &  0.80 \nl
 49344 &  14.30 &  0.80 \nl
 49347 &  14.40 &  0.90 \nl
 49348 &  14.20 &  0.70 \nl
 49349 &  14.60 &  1.00 \nl
 49350 &  15.40 &  1.10 \nl
 49351 &  15.30 &  1.00 \nl
 49352 &  15.10 &  0.90 \nl
 49353 &  15.20 &  0.80 \nl
 49354 &  15.20 &  1.00 \nl
 49355 &  15.10 &  0.80 \nl
 49356 &  15.40 &  1.20 \nl
 49357 &  15.70 &  1.10 \nl
 49358 &  15.50 &  1.00 \nl
 49359 &  15.10 &  0.90 \nl
 49360 &  15.20 &  1.00 \nl
 49361 &  14.90 &  1.00 \nl
 49409 &  14.00 &  0.80 \nl
 49413 &  13.40 &  0.90 \nl
 49422 &  15.10 &  0.70 \nl
 49450 &  13.70 &  1.00 \nl
 49453 &  17.00 &  1.00 \nl
 49474 &  14.80 &  1.50 \nl
\enddata
\end{planotable}

\begin{planotable}{lcc}
\tablewidth{0pt}
\tablenum{4f}
\tablecaption{3.0 mm fluxes of \3c (Tornikoski, Ter\"asranta, Valtaoja). }
\tablehead{ 
\colhead{MJD}&\colhead{Flux}&\colhead{$\sigma$}\\
\colhead{   }&\colhead{(Jy)}&\colhead{(Jy)}
}
\startdata
 47896 &  16.96 &  0.69\nl
 47931 &  14.89 &  0.60\nl
 47937 &  15.61 &  0.63\nl
 47951 &  15.78 &  0.63\nl
 47966 &  14.53 &  0.58\nl
 47981 &  12.84 &  0.52\nl
 47988 &  13.26 &  0.53\nl
 47989 &  13.47 &  0.54\nl
 47994 &  14.11 &  0.57\nl
 47995 &  13.94 &  0.56\nl
 47996 &  13.85 &  0.56\nl
 47997 &  14.27 &  0.57\nl
 47998 &  14.44 &  0.58\nl
 48004 &  14.23 &  0.57\nl
 48011 &  15.26 &  0.61\nl
 48022 &  17.76 &  0.71\nl
 48053 &  28.11 &  1.14\nl
 48062 &  29.61 &  1.19\nl
 48071 &  32.38 &  1.30\nl
 48104 &  33.49 &  1.35\nl
 48231 &  29.17 &  1.17\nl
 48256 &  25.31 &  1.08\nl
 48314 &  32.23 &  1.29\nl
 48351 &  36.22 &  1.46\nl
 48430 &  33.86 &  1.36\nl
 48455 &  34.93 &  1.40\nl
 48469 &  36.10 &  1.50\nl
 48591 &  29.98 &  1.20\nl
 48592 &  29.70 &  1.19\nl
 48614 &  27.13 &  1.09\nl
 48615 &  26.61 &  1.07\nl
 48664 &  29.47 &  1.19\nl
 48665 &  28.03 &  1.12\nl
 48726 &  28.10 &  1.13\nl
 48785 &  25.38 &  1.02\nl
 48857 &  25.21 &  1.01\nl
 48987 &  18.28 &  0.75\nl
 49043 &  21.80 &  0.88\nl
 49078 &  26.30 &  1.05\nl
 49139 &  28.57 &  1.14\nl
 49155 &  23.43 &  0.94\nl
 49323 &  23.13 &  0.93\nl
 49326 &  22.51 &  0.90\nl
 49326 &  23.64 &  0.95\nl
 49350 &  23.37 &  0.94\nl
 49384 &  20.70 &  0.83\nl
 49384 &  22.41 &  0.90\nl
 49385 &  20.62 &  0.83\nl
 49385 &  23.15 &  0.93\nl
 49466 &  19.91 &  0.81\nl
 49467 &  20.14 &  0.81\nl
 49504 &  20.25 &  0.82\nl
 49528 &  21.97 &  0.88\nl
 49565 &  21.38 &  0.86\nl
 49698 &  22.03 &  0.89\nl
 49699 &  21.01 &  0.96\nl
 49749 &  18.10 &  0.73\nl
 49776 &  16.50 &  0.66\nl
 49776 &  16.94 &  0.68\nl
\enddata
\end{planotable}

\begin{planotable}{lcccccccccccccccc}
\scriptsize
\tablecolumns{9}
\tablewidth{0pt}
\tablenum{5}
\tablecaption{Summary of IR and optical observations of \3c (Courvoisier, Marscher, Robson, Wagner).}
\tablehead{  
\colhead{}&
\multicolumn{2}{c}{ U }&
\multicolumn{2}{c}{ B }&
\multicolumn{2}{c}{ V }&
\multicolumn{2}{c}{ R }&
\multicolumn{2}{c}{ I }&
\multicolumn{2}{c}{ J }& 
\multicolumn{2}{c}{ H }&
\multicolumn{2}{c}{ K } \\
\colhead{MJD}&
\colhead{ Flux  }&\colhead{ $\sigma$ }&
\colhead{ Flux  }&\colhead{ $\sigma$ }&
\colhead{ Flux  }&\colhead{ $\sigma$ }&
\colhead{ Flux  }&\colhead{ $\sigma$ }&
\colhead{ Flux  }&\colhead{ $\sigma$ }&
\colhead{ Flux  }&\colhead{ $\sigma$ }&
\colhead{ Flux  }&\colhead{ $\sigma$ }&
\colhead{ Flux  }&\colhead{ $\sigma$ }\\
\colhead{   }&
\colhead{ (mJy)	}&\colhead{  (mJy) }&
\colhead{ (mJy)	}&\colhead{  (mJy) }&
\colhead{ (mJy)	}&\colhead{  (mJy) }&
\colhead{ (mJy)	}&\colhead{  (mJy) }&
\colhead{ (mJy)	}&\colhead{  (mJy) }&
\colhead{ (mJy)	}&\colhead{  (mJy) }&
\colhead{ (mJy)	}&\colhead{  (mJy) }&
\colhead{ (mJy)	}&\colhead{  (mJy) }
}
\startdata
 49078 & \nodata & \nodata & 30.34 & 1.37 & 30.76 & 1.38 & 29.24 & 1.32 & 34.51 & 1.55 & \nodata & \nodata & \nodata & \nodata & \nodata & \nodata  \nl
 49304 & \nodata & \nodata & \nodata & \nodata & \nodata & \nodata & 26.67 & 0.49 & \nodata & \nodata & \nodata & \nodata & \nodata & \nodata & \nodata & \nodata  \nl
 49310 & \nodata & \nodata & \nodata & \nodata & \nodata & \nodata & 26.92 & 0.49 & \nodata & \nodata & \nodata & \nodata & \nodata & \nodata & \nodata & \nodata  \nl
 49330 & \nodata & \nodata & \nodata & \nodata & \nodata & \nodata & \nodata & \nodata & \nodata & \nodata & 38.02 & 1.71 & 50.82 & 2.29 & 84.33 & 3.80  \nl
 49331 & \nodata & \nodata & \nodata & \nodata & \nodata & \nodata & \nodata & \nodata & \nodata & \nodata & 41.30 & 1.86 & 52.24 & 2.35 & 88.31 & 4.75  \nl
 49372 & \nodata & \nodata & 30.62 & 1.38 & 31.05 & 1.40 & 29.24 & 1.32 & 26.92 & 1.21 & \nodata & \nodata & \nodata & \nodata & \nodata & \nodata  \nl
 49386 & 31.14 & 0.28 & 30.37 & 0.20 & 33.18 & 0.19 & \nodata & \nodata & \nodata & \nodata & \nodata & \nodata & \nodata & \nodata & \nodata & \nodata  \nl
 49398 & \nodata & \nodata & \nodata & \nodata & \nodata & \nodata & 27.42 & 0.50 & \nodata & \nodata & \nodata & \nodata & \nodata & \nodata & \nodata & \nodata  \nl
 49399 & 31.98 & 0.28 & 31.48 & 0.20 & 33.89 & 0.19 & \nodata & \nodata & \nodata & \nodata & \nodata & \nodata & \nodata & \nodata & \nodata & \nodata  \nl
 49400 & 32.42 & 0.28 & 31.66 & 0.20 & 33.64 & 0.19 & 27.67 & 0.51 & \nodata & \nodata & \nodata & \nodata & \nodata & \nodata & \nodata & \nodata  \nl
 49414 & 33.12 & 0.28 & 32.10 & 0.20 & 34.65 & 0.19 & \nodata & \nodata & \nodata & \nodata & \nodata & \nodata & \nodata & \nodata & \nodata & \nodata  \nl
\enddata
\end{planotable} 
\begin{planotable}{lcccccc}
\tablenum{6}
\tablecolumns{7}
\tablewidth{0pt}
\tablecaption{Summary of IUE/EUVE observations of \3c (Kafatos).
UV-fluxes are corrected for interstellar absorption using a value for
E(B-V) = 0.03 and the reddening law of Seaton (1979).}
\tablehead{ 
\colhead{MJD}&
\multicolumn{2}{c}{ LWP(1.07\c10{15}~Hz)} &
\multicolumn{2}{c}{ SWP(2.14\c10{15}~Hz)}&
\multicolumn{2}{c}{ DSS(2.99\c10{16}~Hz)}\\
\colhead{}&\colhead{Flux (mJy)}&\colhead{$\sigma$ (mJy)}&\colhead{Flux (mJy)}&
\colhead{ $\sigma$ (mJy)}&\colhead{ Flux (mJy)}&\colhead{ $\sigma$ (mJy)}
}
\startdata
 49360  &  26.30 &  0.99 &  13.93 & 0.48 & \nodata & \nodata \nl
 49523  &  27.56 &  0.99 &  16.36 & 0.48 & \nodata & \nodata \nl
 49360-49366  & \nodata & \nodata & \nodata & \nodata & 0.319  & 0.064 \nl
 49720-49729  &  24.84 &  0.76 &  13.46 & 0.36 & 0.195 & 0.039 \nl
\enddata
\end{planotable}

\begin{planotable}{lcccccc}
\tablewidth{0pt}
\tablenum{7}
\tablecaption{ASCA spectra and fluxes of \3c in December 1993 (Makino, Kii).}
\tablehead{ 
\colhead{Date  }&\colhead{Focal Plane }&\colhead{mean frequency}&\colhead{Flux density }&\colhead{Flux (2-10 keV)}&\colhead{Photon Indices }\\  
\colhead{(MJD)}&\colhead{ Instruments }&\colhead{ (\10{18}Hz) }&\colhead{($\mu$Jy) }&\colhead{(\10{-10}erg/cm$^2$s)}&\colhead{(90\% error area)} 
}
\startdata
Dec 16&  SIS-S0&1.058 & 9.84& 1.72&  1.60 (1.59-1.62)\nl
(49337)& SIS-S1&1.056 & 9.51& 1.66&  1.61 (1.59-1.62)\nl
&        GIS-S2&1.061 & 9.30& 1.63&  1.59 (1.57-1.60)\nl
&        GIS-S3&1.058 & 9.26& 1.62&  1.60 (1.58-1.61)\nl
       	                   			        
Dec 19&  SIS-S0&1.065 & 8.63& 1.52&  1.57 (1.56-1.58)\nl
(49340)& SIS-S1&1.061 & 8.39& 1.47&  1.59 (1.58-1.60)\nl
&        GIS-S2&1.067 & 7.59& 1.34&  1.56 (1.55-1.58)\nl
&        GIS-S3&1.067 & 7.70& 1.36&  1.56 (1.55-1.57)\nl
       	                   			        
Dec 20&  SIS-S0&1.058 & 8.12& 1.42&  1.60 (1.59-1.62)\nl
(49341)& SIS-S1&1.061 & 7.93& 1.39&  1.59 (1.57-1.61)\nl
&        GIS-S2&1.061 & 7.19& 1.26&  1.59 (1.57-1.61)\nl
&        GIS-S3&1.063 & 7.17& 1.26&  1.58 (1.56-1.59)\nl
       	        	        		        
Dec 23&  SIS-S0&1.065 & 8.40& 1.48&  1.57 (1.56-1.59)\nl
(49344)& SIS-S1&1.067 & 8.33& 1.47&  1.56 (1.55-1.58)\nl
&        GIS-S2&1.069 & 7.57& 1.34&  1.55 (1.53-1.57)\nl
&        GIS-S3&1.069 & 7.57& 1.34&  1.55 (1.53-1.56)\nl
       	        	        		        
Dec 27&  SIS-S0&1.074 &10.30& 1.83&  1.53 (1.52-1.55)\nl
(49348)& SIS-S1&1.079 &10.42& 1.86&  1.51 (1.49-1.52)\nl
&        GIS-S2&1.081 & 9.17& 1.64&  1.50 (1.49-1.52)\nl
&        GIS-S3&1.079 & 9.19& 1.64&  1.51 (1.49-1.53)\nl
\enddata
\end{planotable}

\begin{planotable}{lcc}
\tablewidth{200pt}
\tablenum{8}
\tablecaption{OSSE spectrum of \3c during MJD 49341-49348 (Johnson).}
\tablehead{ 
\colhead{mean frequency }&\colhead{ Flux  }&\colhead{ $\sigma$  }\\
\colhead{(Hz)}&\colhead{ ($\mu$Jy)}&\colhead{  ($\mu$Jy)}
}
\startdata
 2.1\c10{19}  &  1.40 & 0.14 \nl
 5.1\c10{19}  &  0.76 & 0.16 \nl
 1.2\c10{20}  &  0.13 & 0.19 \nl
 2.6\c10{20}  &  0.48 & 0.35 \nl
 5.9\c10{20}  &  0.00 & 0.22 \nl
 1.5\c10{21}  &  0.11 & 0.10 \nl
\enddata
\end{planotable}

\begin{planotable}{lcc}
\tablewidth{200pt}
\tablenum{9}
\tablecaption{COMPTEL spectrum of \3c during MJD 49279-49355 (Collmar).}
\tablehead{ 
\colhead{mean frequency }&\colhead{ Flux  }&\colhead{ $\sigma$  }\\
\colhead{(Hz)}&\colhead{ ($\mu$Jy)}&\colhead{  ($\mu$Jy)}
}
\startdata
 2.1\c10{20}  &  0.12       &  0.077   \nl
 4.0\c10{20}  &  0.097      &  0.023   \nl
 1.2\c10{21}  &  0.036      &  0.0082  \nl
 4.0\c10{21}  &  $<$ 0.0078 &          \nl
\enddata
\end{planotable}

\end{document}